\documentclass[11pt]{article}

\usepackage{amssymb}
\usepackage{citesort}

\begin{document}
\begin{titlepage}
\thispagestyle{empty}

\begin{flushright}
                hep-th/0409258 \\
                TPJU -- 14/2004\\
                01 09/ 2004 IFT UWr
\end{flushright}
\bigskip

\begin{center}

{\LARGE\bf\sf Analytic continuation formulae \\
\vskip 3mm
for the BPZ conformal block}

\end{center}
\bigskip

\begin{center}
    {\large\bf\sf
    Leszek Hadasz}\footnote{e-mail: hadasz@th.if.uj.edu.pl} \\
\vskip 1mm
    M. Smoluchowski Institute of Physics, \\
    Jagiellonian University,
    Reymonta 4, 30-059 Krak\'ow, Poland \\
\vskip 5mm
    {\large\bf\sf
    Zbigniew Jask\'{o}lski}\footnote{e-mail: jask@ift.uni.wroc.pl
}\\
{\large\bf\sf
    Marcin Pi\c{a}tek}\footnote{e-mail: piatek@ift.uni.wroc.pl
}\\
\vskip 1mm
   Institute of Theoretical Physics\\
University of Wroc{\l}aw\\
 pl. M. Borna, 950-204 Wroc{\l}aw
 Poland
\end{center}

\vskip .5cm

\begin{abstract}
Using the techniques developed by Ponsot and Teschner \cite{Ponsot:1999uf,Ponsot:2000mt,Ponsot:2001ng}
and Teschner \cite{Teschner:2001rv,Teschner:2003en}
we derive the formulae for analytic continuation of the general
4-point conformal block.
\end{abstract}

\end{titlepage}
\newpage
\vskip 1cm
\section{Introduction}
The operator product
expansion of primary fields
in the standard CFT \cite{Belavin:1984vu}
can be written as
\begin{equation}
\phi_{\Delta_1\bar \Delta_1}(z,\bar {z})\phi_{\Delta_2\bar \Delta_2}(0,0)=\sum_{p}
C_{12p}
\,z^{\Delta_{p}-\Delta_{1}-\Delta_{2}}
\overline{z}^{\,\,\bar {\Delta}_{p}-\bar{\Delta}_{1}-\bar{\Delta}_{2}}
\Psi_{12p}(z,\bar z),
\end{equation}
where for each $p$ the descendent field $\Psi_{12p}(z,\bar z)$ is
uniquely determined by the conformal invariance. Acting on the
vacuum $|\,0\, \rangle $ it generates a state $ \Psi_{12p}(z,\bar
z) |\,0\,\rangle =|\psi_{12p}(z)\rangle\otimes |\bar
\psi_{12p}(\bar{z})\rangle $ in the tensor product ${\cal
V}_{\Delta_p}\otimes {\cal V}_{\bar \Delta_p}$ of the Verma
modules with the the highest weights $\Delta_p$ and $\bar
\Delta_p$ respectively. The $z$ dependence of each component is
uniquely determined by the conformal invariance. In the ``left''
(holomorphic) sector one has
\begin{eqnarray*}
|\psi_{12p}(z)\rangle &=&\nu_p + \sum\limits_{n=1}^\infty z^n
\beta^{\,n}_{ \Delta_{p}}\!\left[\,^{\Delta_2}_{\Delta_1}\right],
\end{eqnarray*}
where
 $\nu_p$ is the highest weight vector in
${\cal V}_{\Delta_p}$ and $ \beta^{\,n}_{c,
\Delta_{p}}\!\left[\,^{\Delta_2}_{\Delta_1}\right] \in {\cal
V}_{\Delta_p}^n\subset {\cal V}_{\Delta_p}$. In the $n-$th level
subspace ${\cal V}^{\;n}_{\Delta_p}$ we shall use  a standard
basis consisting of vectors of the form
\begin{equation}
\label{basis}
\nu^n_{p,I}
\; = \;
L_{-I}\nu_p
\; = \;
L_{-i_{k}}\ldots L_{-i_{2}}L_{-i_{1}} \nu_p\ ,
\end{equation}
where $I=\lbrace
i_{k},\ldots ,i_{1}\rbrace$ is an ordered ( $i_{k}\geq \ldots\geq
i_{1}\geq 1$) sequence of positive integers of the length $|I|\equiv i_{k}+\ldots+i_{1}=n$.

The conformal Ward identity for the
3-point function implies  the equations
$$
L_i\, \beta^{\,n}_{\Delta}\!\left[\,^{\Delta_2}_{\Delta_1}\right]
\; = \;
(\Delta_p + i\Delta_1 -\Delta_2 + n -i)
\beta^{\,n-i}_{ \Delta}\!\left[\,^{\Delta_2}_{\Delta_1}\right],
$$
which in the basis (\ref{basis})
take the form
\begin{equation}
\label{betas}
\sum\limits_{|J|=n} \Big[ G_{\Delta}\Big]_{IJ}
\beta^{\,n}_{\Delta}\!\left[\,^{\Delta_2}_{\Delta_1}\right]^J
\; = \;
\gamma_{ \Delta}\!\left[\,^{\Delta_2}_{\Delta_1}\right]_I,
\end{equation}
where $ \Big[ G_{\Delta}\Big]_{IJ}=\langle \nu_I,\nu_J \rangle$ is the Gram matrix
of the standard symmetric bilinear form in ${\cal V}_{\Delta_p}$ and
\begin{eqnarray}
\label{gamma}
\gamma_{ \Delta}\!\left[\,^{\Delta_2}_{\Delta_1}\right]_I\!\! &=&\!\!
(\Delta +i_k\Delta_1 -\Delta_2 + i_{k-1} \!+\!\dots \!+\!i_1)\times \dots\\
\nonumber
&\dots &\times(\Delta +i_2\Delta_1 -\Delta_2 +i_1) (\Delta +i_1\Delta_1 -\Delta_2 ).
\end{eqnarray}
For all  values of the variables $c$ and $\Delta$ for which the Gram matrices are invertible
the equations (\ref{betas}) admit  unique solutions
$$
\beta^{\,n}_{\Delta}\!\left[\,^{\Delta_2}_{\Delta_1}\right]^I \; =\;
\sum\limits_{|J|=n} \Big[ G_{\Delta}\Big]^{IJ} \gamma_{
\Delta}\!\left[\,^{\Delta_2}_{\Delta_1}\right]_J.
$$
In this range
the 4-point conformal block is defined as a formal power series  \cite{Belavin:1984vu}
\begin{eqnarray}
\label{block}
{\cal F}_{\Delta}\!\left[_{\Delta_{4}\;\Delta_{1}}^{\Delta_{3}\;\Delta_{2}}\right]\!(\,z)
&=&
z^{\Delta-\Delta_{2}-\Delta_{1}}\left( 1 +
\sum_{n=1}^\infty z^{\;n}
{\cal F}^{\,n}_{\Delta}\!\left[_{\Delta_{4}\;\Delta_{1}}^{\Delta_{3}\;\Delta_{2}}\right] \right),
\\
{\cal F}^{\,n}_{\Delta}\!\left[_{\Delta_{4}\;\Delta_{1}}^{\Delta_{3}\;\Delta_{2}}\right]
&=&
\sum\limits_{|I|=|J|=n}
\gamma_{\Delta}\!\left[\,^{\Delta_3}_{\Delta_4}\right]_I
\Big[ G_{\Delta}\Big]^{IJ}
\gamma_{\Delta}\!\left[\,^{\Delta_2}_{\Delta_1}\right]_J.
\nonumber
\end{eqnarray}
One of the long standing open problems has been the calculation of the radius
of convergence of this series. The basic difficulty is that
no closed formula for the coefficients is known,
what makes a direct analysis prohibitively difficult.
As
the dimension of ${\cal V}^n_\Delta$ grows rapidly with $n$
also numerical calculations  by inverting
the Gram matrices became very laborious for higher orders.

A more efficient method based on a recurrence relation for the
coefficients was developed by Al.~B.~Zamolodchikov
\cite{Zamolodchikov:ie}. It should be stressed however that the
analytic properties of the conformal block with respect to its
parameters which are crucial for this method are derived under the
assumption that the radius of convergence is nonzero. Also some
conjectured formulae of this approach are still to be rigorously
proved.

The commonly accepted hypothesis justified by all special cases
where the conformal block can be explicitly calculated is that the
series (\ref{block}) converges for all $|z|< 1$.
It has been recently suggested \cite{Teschner:2001rv}
to apply the free field representation
of the chiral vertex operator \cite{Teschner:2001rv,Teschner:2003en}
to solve this problem.
We assume in the present paper that the  radius of convergence is 1.

Another hypothesis supported by all known examples  concerns an
analytic continuation of the function defined by the series (\ref{block}).
It states that the only singularities of the conformal block with
respect to the $z$ variable are branching points (in general of a transcendental kind)
at $0$, $1$, and $\infty$ \cite{ZZ}. This in particular means
that the conformal block is a single-valued analytic function on the
universal covering of a 3-punctured Riemann sphere and can
be expressed  by a power series convergent in the entire domain
of its analyticity \cite{Zam}. A recurrence relation for
calculating coefficients of this so called $q$-expansion
\cite{Zam} provides an efficient method
for numerical analysis of conformal block and can be applied for
testing the conformal bootstrap equations
\cite{Zamolodchikov:1995aa,Runkel:2001ng}.

In the RCFT models where the number of conformal blocks is finite
the problem of analytic continuation is essentially
equivalent to the problem of calculating
the monodromy matrices relating conformal blocks in different
channels \cite{Lewellen:1988sv,Blok:1988ku,Moore:1988qv}. A thorough analysis of the consistency conditions
such matrices have to satisfy was done by Moore and Seiberg in terms
of the braiding and fusion relations of
chiral vertex operators \cite{Moore:1988qv}.

The Moore--Seiberg formalism
suitably generalized to the case of continuous spectrum
was recently applied by Ponsot and Teschner
to derive a system of functional equations
for the braiding and fusion matrices of the Liouville theory.
They  also constructed explicit solutions to these equations
by means of the representation theory of
${\cal U}_q({\rm sl}(2,\mathbb{R}))$ \cite{Ponsot:1999uf,Ponsot:2000mt}.
The exact form of the braiding and fusion matrices
can be also
derived by direct calculations of the exchange relation of
chiral vertex operators in the
 free field representation \cite{Teschner:2001rv,Teschner:2003en}
 (see also \cite{Gervais:1993fh} for an earlier construction).

In the present
note we use the results of \cite{Ponsot:1999uf,Ponsot:2000mt,Ponsot:2001ng,Teschner:2001rv,Teschner:2003en,Ponsot:2003ju}
to derive the formulae for the analytic continuation of
the BPZ conformal block. To this end a choice of cuts for the function (\ref{block})
has to be made. We pick the cut starting at the origin to run along the negative real axis,
and the one starting at 1 to run along the positive real axis.
With this choice the formulae read:
\begin{eqnarray}
\label{s-u}
{\cal F}_{\Delta_s}\!\left[_{\Delta_{4}\;\Delta_{1}}^{\Delta_{3}\;\Delta_{2}}\right]\!\left(z\right)
&=& z^{-2\Delta_2}
\\
\nonumber
&&\hspace{-50pt}\times\;
{1\over 2i}\!\!\!
\int\limits_{{Q\over 2} +i \mathbb{R}}\!\!\!\! d\alpha_u
\;{\rm e}^{-\epsilon\pi i(\Delta_1+\Delta_4-\Delta_s-\Delta_u)}
B_{\alpha_s\,\alpha_u}\!\Big[ \,{_{\alpha_{4}\;\alpha_{1}}^{\alpha_{3}\;\alpha_{2}}} \Big]
{\cal F}_{\Delta_u}\!\left[ _{\Delta_{1}\;\Delta_{4}}^{\Delta_{3}\;\Delta_{2}}\right]\!\left({1\over z}\right),
\\[20pt]
\label{euler}
{\cal F}_{\Delta_s}\!\left[_{\Delta_{4}\;\Delta_{1}}^{\Delta_{3}\;\Delta_{2}}\right]\!\left(z\right)
&=&
(1-z)^{\Delta_4-\Delta_3-\Delta_2-\Delta_1}
\;{\rm e}^{- \epsilon\pi i (\Delta_s-\Delta_2-\Delta_1)}
{\cal F}_{\Delta_s}\!\left[_{\Delta_{4}\;\Delta_{2}}^{\Delta_{3}\;\Delta_{1}}\right]\!\left({z\over z-1}\right),
\\[20pt]
\label{s-t}
{\cal F}_{\Delta_s}\!\left[_{\Delta_{4}\;\Delta_{1}}^{\Delta_{3}\;\Delta_{2}}\right]\!\left(z\right)
&=&z^{\Delta_4-\Delta_3-\Delta_2-\Delta_1}
\\
\nonumber
&&\hspace{-50pt} \times\;
{1\over 2i}\!\!\!\int\limits_{{Q\over 2} +i \mathbb{R}}\!\!\!\! d\alpha_t
\;{\rm e}^{\epsilon\pi i(\Delta_t-\Delta_3-\Delta_2)}
B_{\alpha_s\,\alpha_t}\!\Big[ \,{_{\alpha_{4}\;\alpha_{2}}^{\alpha_{3}\;\alpha_{1}}} \Big]
{\cal F}_{\Delta_t}\!\left[ _{\Delta_{2}\;\Delta_{4}}^{\Delta_{3}\;\Delta_{1}}\right]\!\left(1-{1\over z}\right),
\\[20pt]
\label{u-t}
{\cal F}_{\Delta_u}\!\left[_{\Delta_{4}\;\Delta_{1}}^{\Delta_{3}\;\Delta_{2}}\right]\!\left(z\right)
&=&
{1\over 2i}\!\!\!\int\limits_{{Q\over 2} +i \mathbb{R}}\!\!\!\! d\alpha_t
\;
B_{\alpha_u\,\alpha_t}\!\Big[ \,{_{\alpha_{4}\;\alpha_{2}}^{\alpha_{3}\;\alpha_{1}}} \Big]
{\cal F}_{\Delta_t}\!\left[_{\Delta_{4}\;\Delta_{3}}^{\Delta_{1}\;\Delta_{2}}\right]\!\left(1-z\right),
\end{eqnarray}
where $\epsilon = +1$ if  $\arg z>0$,  $\epsilon = -1$ if  $\arg z<0$,  and we have used the standard
parameterizations of the central charge and the conformal weights:
\begin{eqnarray*}
c&=&1+6Q^2,
\hskip 5mm
Q\;=\;b+{1\over b},
\\
\Delta_j &=&\Delta(\alpha_j)\;=\;\alpha_j(Q-\alpha_j),
\hskip 5mm
j\;=\;s,\, t,\, u,\, 1,\dots,4.
\end{eqnarray*}
The first two equations are straightforward consequences of the braiding
relation derived in \cite{Teschner:2001rv,Teschner:2003en}.
The next two are less obvious and
their derivation is our main objective in this paper.

The braiding matrix calculated
in the free field representation from the exchange relation of chiral
vertex operators is given by \cite{Teschner:2001rv,Teschner:2003en}:
\begin{eqnarray}
\label{BE}
&& \hspace{-20pt}
B^\epsilon_{\alpha_s\,\alpha_u}\!\Big[ \,{_{\alpha_{4}\;\alpha_{1}}^{\alpha_{3}\;\alpha_{2}}} \Big]
\;=\; {\rm e}^{\epsilon\pi i(\Delta_1+\Delta_4-\Delta_s-\Delta_u)}
B_{\alpha_s\,\alpha_u}\!\Big[ \,{_{\alpha_{4}\;\alpha_{1}}^{\alpha_{3}\;\alpha_{2}}} \Big],
\\[15pt]
\label{B}
&& \hspace{-20pt} B_{\alpha_s\,\alpha_u}\!\Big[ \,{_{\alpha_{4}\;\alpha_{1}}^{\alpha_{3}\;\alpha_{2}}} \Big]
\;=\;
\\
&&\nonumber
{\Gamma_b(\bar\alpha_4+\bar\alpha_2-\alpha_u)
\Gamma_b(\alpha_4+\bar\alpha_2 -\alpha_u)
\Gamma_b(\bar\alpha_4-\alpha_2+ \alpha_u)
\Gamma_b(\alpha_4-\alpha_2+\alpha_u)
\over
\Gamma_b(\bar\alpha_4+\bar\alpha_3-\alpha_s)
\Gamma_b(\alpha_4+\bar\alpha_3 -\alpha_s)
\Gamma_b(\bar\alpha_4-\alpha_3+ \alpha_s)
\Gamma_b(\alpha_4-\alpha_3+\alpha_s)
}
\\
\nonumber
&\times&
{\Gamma_b(\bar\alpha_1+\bar\alpha_3 -\alpha_u)
\Gamma_b(\alpha_1+\bar\alpha_3-\alpha_u)
\Gamma_b(\bar\alpha_1-\alpha_3+\alpha_u)
\Gamma_b(\alpha_1-\alpha_3+\alpha_u)
\over
\Gamma_b(\bar\alpha_1+\bar\alpha_2 -\alpha_s)
\Gamma_b(\alpha_1+\bar\alpha_2-\alpha_s)
\Gamma_b(\bar\alpha_1-\alpha_2+\alpha_s)
\Gamma_b(\alpha_1-\alpha_2+\alpha_s)}
\\
\nonumber
&\times &
{\Gamma_b(2\alpha_s)\Gamma_b(2\bar\alpha_s)\over
\Gamma_b(\bar\alpha_u-\alpha_u)\Gamma_b(\alpha_u-\bar\alpha_u)}
\\
\nonumber
&\times &
{1\over i}\int\limits_{i\mathbb{R}}\!\!dt\;
{S_b(\bar\alpha_1+t)
S_b(\alpha_1+t)
S_b(\bar\alpha_4-\alpha_3+\alpha_2+t)
S_b(\alpha_4-\alpha_3+\alpha_2+t)
\over
S_b(\bar\alpha_s+\alpha_2+t)
S_b(\alpha_s+\alpha_2+t)
S_b(\bar\alpha_u+\bar\alpha_3+t)
S_b(\alpha_u+\bar\alpha_3+t)
},
\end{eqnarray}
where $\Gamma_b(z)$ is the Barnes double gamma function,
$S_b(z) \equiv {\Gamma_b(z)\over\Gamma_b(Q-z)}$\footnote{
Some of the properties of these special functions are collected in Appendix A},
and we have use the abbreviated notation
$\bar\alpha \equiv Q-\alpha$.
For $\alpha_j\in {Q\over 2}+i\mathbb{R},\ j=s,\,t,\,u,\,1,\dots,4$
corresponding to the spectrum of the Liouville theory  the integrand in (\ref{s-t})
has simple poles on the imaginary axis. The contour of integration is
located to the left of all these poles.

The matrix
$
B_{\alpha_s\,\alpha_u}\!
\Big[ \,{_{\alpha_{4}\;\alpha_{1}}^{\alpha_{3}\;\alpha_{2}}} \Big]
$
is related to the fusion matrix  for the normalized
chiral vertex operators \cite{Ponsot:2001ng}
by the simple exchange $\alpha_1 \leftrightarrow \alpha_2$ of its parameters:
$$
B_{\alpha_s\,\alpha_u}\!
\Big[ \,{_{\alpha_{4}\;\alpha_{1}}^{\alpha_{3}\;\alpha_{2}}} \Big]
\;=\;F_{\alpha_s\,\alpha_u}\!
\Big[ \,{_{\bar\alpha_{3}\;\alpha_{1}}^{\bar\alpha_{4}\;\alpha_{2}}} \Big]
\;=\;
F_{\alpha_s\,\alpha_u}\!
\Big[ \,{_{\alpha_{4}\;\alpha_{2}}^{\alpha_{3}\;\alpha_{1}}} \Big]
$$
It is symmetric with respect to the exchange of column and rows,
as well as with respect to the change $\alpha_j \to Q-\alpha_j$ in each
$\alpha_j$ separately \cite{Ponsot:1999uf,Ponsot:2000mt,Ponsot:2001ng,Ponsot:2003ju}.
The latter property means that the matrices
(\ref{BE}), (\ref{B})  depend only on conformal weights.
An explicit derivation of
these properties along the line mentioned in \cite{Ponsot:2003ju} is presented in Appendix B.

 Both the formulae (\ref{s-u}) -- (\ref{u-t})
and the expressions (\ref{BE}), (\ref{B}) admit an analytic   continuation
to generic values of $\alpha_j$. We present  a special example of such continuation to a degenerate weight
in Section 3.

Finally let us note that possible applications of the analytic continuation formulae
goes beyond the Liouville theory.
They are powerful tools not only for analyzing general properties of the conformal block like
the conjectured analytic structure  but also for explicit calculations.

\section{Derivation}

We shall work with the matrix elements of the chiral vertex operators  \cite{Moore:1988qv}
rather than
with the operators themselves.
For a given triple $\Delta_1,\Delta_2, \Delta_3$ of conformal weights
we define the matrix element of a single chiral vertex operator as a trilinear map
$$
\rho{^{\Delta_3}_\infty}{^{\Delta_2}_{\:z}}{^{\Delta_1}_{\;0}} :
{\cal V}_{\Delta_3}\times {\cal V}_{\Delta_2} \times {\cal V}_{\Delta_1}
\to
\mathbb{C}
$$
satisfying  the following conditions\footnote{
These are just the well known conditions for the chiral vertex operator
written in terms of its matrix elements.}
 \cite{Teschner:2001rv}:
\begin{eqnarray}
\label{lcommutator}
\rho{^{\Delta_3}_\infty}{^{\Delta_2}_{\:z}}{^{\Delta_1}_{\;0}}
(L_{-n} \xi_3,\nu_2,\xi_1)
&=&\rho{^{\Delta_3}_\infty}{^{\Delta_2}_{\:z}}{^{\Delta_1}_{\;0}}
( \xi_3,\nu_2,L_n\xi_1)
\\
\nonumber
&+&
 z^{n}(z \partial_z + (n+1)\Delta_2 )
 \rho{^{\Delta_3}_\infty}{^{\Delta_2}_{\:z}}{^{\Delta_1}_{\;0}}
(\xi_3,\nu_2,\xi_1),
\\[15pt]
\label{translation}
\rho{^{\Delta_3}_\infty}{^{\Delta_2}_{\:z}}{^{\Delta_1}_{\;0}}
( \xi_3,L_{-1}\xi_2,\xi_1)
 &=&
 \partial_z
 \rho{^{\Delta_3}_\infty}{^{\Delta_2}_{\:z}}{^{\Delta_1}_{\;0}}
( \xi_3,\xi_2,\xi_1),
\\[15pt]
\label{L2a}
\rho{^{\Delta_3}_\infty}{^{\Delta_2}_{\:z}}{^{\Delta_1}_{\;0}}
( \xi_3,L_{n}\xi_2,\xi_1)
&=&
 \sum\limits_{k=0}^{n+1} \left(\,_{\;\;k}^{n+1}\right) (-z)^{k}
\left(\rho{^{\Delta_3}_\infty}{^{\Delta_2}_{\:z}}{^{\Delta_1}_{\;0}}
( L_{k-n}\xi_3,\xi_2,\xi_1)\right.
\\
\nonumber
&&\hspace{70pt}
-\;
\left. \rho{^{\Delta_3}_\infty}{^{\Delta_2}_{\:z}}{^{\Delta_1}_{\;0}}
( \xi_3,\xi_2,L_{n-k}\xi_1) \right),
\hskip 5mm n>-1,
\\[15pt]
\label{L2b}
\rho{^{\Delta_3}_\infty}{^{\Delta_2}_{\:z}}{^{\Delta_1}_{\;0}}
( \xi_3,L_{-n}\xi_2,\xi_1)
&=&
 \sum\limits_{k=0}^{\infty} \left(\,_{\;\;n-2}^{n-2+k}\right)
z^{k}
\rho{^{\Delta_3}_\infty}{^{\Delta_2}_{\:z}}{^{\Delta_1}_{\;0}}
(L_{n+k} \xi_3,\xi_2,\xi_1)
\\
\nonumber
&&
\hspace{-40pt}
+\;(-1)^n \sum\limits_{k=0}^{\infty} \left(\,_{\;\;n-2}^{n-2+k}\right)
z^{-n+1-k}
\rho{^{\Delta_3}_\infty}{^{\Delta_2}_{\:z}}{^{\Delta_1}_{\;0}}
( \xi_3,\xi_2, L_{k-1}\xi_1),
\hskip 5mm n>1,
\\[15pt]
\label{normalization}
\rho{^{\Delta_3}_\infty}{^{\Delta_2}_{\:z}}{^{\Delta_1}_{\;0}}
( \nu_3,\nu_2,\nu_1)
&=&
z^{\Delta_3- \Delta_2- \Delta_1 },
\end{eqnarray}
where $\nu_i$ is the highest weight state in ${\cal V}_{\Delta_i}$ ($i=1,2,3$).

 The form
$\displaystyle \rho{^{\Delta_3}_\infty}{^{\Delta_2}_{\:z}}{^{\Delta_1}_{\;0}}$
is uniquely determined by the properties above.
In particular,
for $L_0$-eingenstates
$
L_0|\xi_i\rangle  = \Delta_i (\xi_i)|\xi_i\rangle,\;i=1,2,3,
$
one has:
\begin{equation}
\label{z_dep}
\rho{^{\Delta_3}_\infty}{^{\Delta_2}_{\:z}}{^{\Delta_1}_{\;0}}
(\xi_3,\xi_2,\xi_1) \; = \;
z^{\Delta_3(\xi_3)- \Delta_2(\xi_2)- \Delta_1(\xi_1) }
\rho{^{\Delta_3}_\infty}{^{\Delta_2}_{\:1}}{^{\Delta_1}_{\;0}}
(\xi_3,\xi_2,\xi_1),
\end{equation}
and
\begin{eqnarray}
\nonumber
\rho{^{\Delta_3}_\infty}{^{\Delta_2}_{\:1}}{^{\Delta_1}_{\;0}}
(\nu_{3,I},\nu_2,\nu_1)
&=& \gamma_{ \Delta_3}\!\left[\,^{\Delta_2}_{\Delta_1}\right]_I,
\\
\label{ggg}
\rho{^{\Delta_3}_\infty}{^{\Delta_2}_{\:1}}{^{\Delta_1}_{\;0}}
(\nu_3,\nu_2,\nu_{1,I})
&=& \gamma_{ \Delta_1}\!\left[\,^{\Delta_2}_{\Delta_3}\right]_I,
\\
\nonumber
\rho{^{\Delta_3}_\infty}{^{\Delta_2}_{\:1}}{^{\Delta_1}_{\;0}}
(\nu_3,\nu_{2,I},\nu_1)
&=& (-1)^{|I|}\gamma_{ \Delta_2}\!\left[\,^{\Delta_1}_{\Delta_3}\right]_I,
\end{eqnarray}
for all vectors  $\nu_{i,I}$ of the form (\ref{basis}).

In order to simplify notation we introduce a graphic representation
of the form $\rho$ and the matrix elements of two
possible compositions of the chiral vertex operators:
\begin{eqnarray*}
\begin{picture}(65,30)(0,5)
%1vertex
\put(0,0){\makebox(15,10)
{$
  \scriptstyle 3
$}}
\put(15,5){\line(1,0){10}}
\put(25,0){\makebox(15,10){$
 \scriptstyle z
$}}
\put(40,5){\line(1,0){10}}
\put(50,0){\makebox(15,10)[c]
{$
  \scriptstyle 1
$}}
\put(32.5,10){\line(0,1){8}}
\put(25,20){\makebox(15,10){$
  \scriptstyle 2
$}}
\end{picture}
&\equiv & \;\rho{^{\Delta_3}_\infty}{^{\Delta_2}_{\:z}}{^{\Delta_1}_{\;0}}
(\nu_3,\nu_2,\nu_1),
\\[15pt]
\begin{picture}(115,30)(0,5)
%2vertex
\put(0,0){\makebox(15,10)
{$
  \scriptstyle 4
$}}
\put(15,5){\line(1,0){10}}
\put(25,0){\makebox(15,10){$
 \scriptstyle z_3
$}}
\put(40,5){\line(1,0){10}}
\put(50,0){\makebox(15,10)[c]
{$
  \scriptstyle s
$}}
\put(65,5){\line(1,0){10}}
\put(75,0){\makebox(15,10){$
 \scriptstyle z_2
$}}
\put(90,5){\line(1,0){10}}
\put(100,0){\makebox(15,10){$
 \scriptstyle  1
$}}
\put(32.5,10){\line(0,1){8}}
\put(25,20){\makebox(15,10){$
  \scriptstyle 3
$}}
\put(82.5,10){\line(0,1){8}}
\put(75,20){\makebox(15,10){$
 \scriptstyle  2
$}}
\end{picture}
&\equiv &
\\[15pt]
&&\hspace{-80pt}
\sum\limits_{I,J}
\rho{^{\Delta_4}_\infty}{^{\Delta_3}_{\,z_3}}{^{\Delta_s}_{\;0}}
(\nu_4\,,\,\nu_3\,,\,\nu_{s,I})
\,[G_{\Delta_s}]^{IJ}
\rho{^{\Delta_s}_\infty}{^{\Delta_2}_{\,z_2}}{^{\Delta_1}_{\;0}}
(\nu_{s,J}\,,\,\nu_2\,,\,\nu_1),
\\
\begin{picture}(90,50)(0,5)
%2vertex
\put(0,0){\makebox(15,10)
{$
  \scriptstyle 4
$}}
\put(15,5){\line(1,0){10}}
\put(25,0){\makebox(15,10){$
 \scriptstyle z_2
$}}
\put(40,5){\line(1,0){10}}
\put(50,0){\makebox(15,10)[c]
{$
  \scriptstyle 1
$}}
\put(32.5,10){\line(0,1){8}}
\put(25,20){\makebox(15,10){$
  \scriptstyle t
$}}
\put(40,25){\line(1,0){10}}
\put(50,20){\makebox(15,10){$
 \scriptstyle z_{32}
$}}
\put(65,25){\line(1,0){10}}
\put(75,20){\makebox(15,10){$
 \scriptstyle  2
$}}
\put(57.5,30){\line(0,1){8}}
\put(50,40){\makebox(15,10){$
 \scriptstyle  3
$}}
\end{picture}
&\equiv &
\\[15pt]
&&\hspace{-80pt}
\sum\limits_{I,J}
\rho{^{\Delta_4}_\infty}{^{\Delta_t}_{\,z_2}}{^{\Delta_1}_{\;0}}
(\nu_4\,,\,\nu_{t,I}\,,\,\nu_1)
\,[G_{\Delta_t}]^{IJ}
\rho{^{\Delta_t}_\infty}{^{\Delta_3}_{z_{32}}}{^{\Delta_2}_{\;0}}
(\nu_{t,J}\,,\,\nu_3\,,\,\nu_2),
\end{eqnarray*}
where the abbreviation $z_{32}=z_3-z_2$ has been used.
Let us note that the compositions are well defined if the
intermediate conformal weights are non-degenerate.

Our basic tool in further considerations is the braiding relation
obtained in  \cite{Teschner:2001rv,Teschner:2003en}:
\begin{eqnarray}
\label{braiding}
\begin{picture}(115,30)(0,5)
%2vertex
\put(0,0){\makebox(15,10)
{$
  \scriptstyle 4
$}}
\put(15,5){\line(1,0){10}}
\put(25,0){\makebox(15,10){$
 \scriptstyle z_3
$}}
\put(40,5){\line(1,0){10}}
\put(50,0){\makebox(15,10)[c]
{$
  \scriptstyle s
$}}
\put(65,5){\line(1,0){10}}
\put(75,0){\makebox(15,10){$
 \scriptstyle z_2
$}}
\put(90,5){\line(1,0){10}}
\put(100,0){\makebox(15,10){$
 \scriptstyle  1
$}}
\put(32.5,10){\line(0,1){8}}
\put(25,20){\makebox(15,10){$
  \scriptstyle 3
$}}
\put(82.5,10){\line(0,1){8}}
\put(75,20){\makebox(15,10){$
 \scriptstyle  2
$}}
\end{picture}
&=&
\\
\nonumber
    &&                \hspace{-90pt}
{1\over 2i}\!\!\!\int\limits_{{Q\over 2}+i \mathbb{R}} d\alpha_u\;
B^{\epsilon_{32}}_{\alpha_s\,\alpha_u}\!\Big[ \,{_{\alpha_{4}\;\alpha_{1}}^{\alpha_{3}\;\alpha_{2}}} \Big]
                                    \hspace{5pt}
\begin{picture}(115,30)(0,5)
%2vertex
\put(0,0){\makebox(15,10)
{$
  \scriptstyle 4
$}}
\put(15,5){\line(1,0){10}}
\put(25,0){\makebox(15,10){$
 \scriptstyle z_2
$}}
\put(40,5){\line(1,0){10}}
\put(50,0){\makebox(15,10)[c]
{$
  \scriptstyle u
$}}
\put(65,5){\line(1,0){10}}
\put(75,0){\makebox(15,10){$
 \scriptstyle z_3
$}}
\put(90,5){\line(1,0){10}}
\put(100,0){\makebox(15,10){$
 \scriptstyle  1
$}}
\put(32.5,10){\line(0,1){8}}
\put(25,20){\makebox(15,10){$
  \scriptstyle 2
$}}
\put(82.5,10){\line(0,1){8}}
\put(75,20){\makebox(15,10){$
 \scriptstyle  3
$}}
\end{picture}
\end{eqnarray}
where $\epsilon_{32} = +1$ if  $\arg z_{32}>0$,  $\epsilon_{32} = -1$ if  $\arg z_{32}<0$.
We shall also need the formulae for the coupling to the
vacuum ($\Delta_0=0$):
\begin{eqnarray}
\label{simple}
\begin{picture}(115,30)(0,5)
%2vertex
\put(0,0){\makebox(15,10)
{$
  \scriptstyle 3
$}}
\put(15,5){\line(1,0){10}}
\put(25,0){\makebox(15,10){$
 \scriptstyle z
$}}
\put(40,5){\line(1,0){10}}
\put(50,0){\makebox(15,10)[c]
{$
  \scriptstyle 1
$}}
\put(65,5){\line(1,0){10}}
\put(75,0){\makebox(15,10){$
 \scriptstyle 0
$}}
\put(90,5){\line(1,0){10}}
\put(100,0){\makebox(15,10){$
 \scriptstyle  0
$}}
\put(32.5,10){\line(0,1){8}}
\put(25,20){\makebox(15,10){$
  \scriptstyle 2
$}}
\put(82.5,10){\line(0,1){8}}
\put(75,20){\makebox(15,10){$
 \scriptstyle  1
$}}
\end{picture}
                                    \hspace{10pt}
                                       &=&
                                    \hspace{10pt}
\begin{picture}(65,30)(0,5)
%1vertex
\put(0,0){\makebox(15,10)
{$
  \scriptstyle 3
$}}
\put(15,5){\line(1,0){10}}
\put(25,0){\makebox(15,10){$
 \scriptstyle z
$}}
\put(40,5){\line(1,0){10}}
\put(50,0){\makebox(15,10)[c]
{$
  \scriptstyle 1
$}}
\put(32.5,10){\line(0,1){8}}
\put(25,20){\makebox(15,10){$
  \scriptstyle 2
$}}
\end{picture}
\\
\label{identity}
\begin{picture}(115,30)(0,5)
%2vertex
\put(0,0){\makebox(15,10)
{$
  \scriptstyle 3
$}}
\put(15,5){\line(1,0){10}}
\put(25,0){\makebox(15,10){$
 \scriptstyle z_2
$}}
\put(40,5){\line(1,0){10}}
\put(50,0){\makebox(15,10)[c]
{$
  \scriptstyle 1
$}}
\put(65,5){\line(1,0){10}}
\put(75,0){\makebox(15,10){$
 \scriptstyle z_1
$}}
\put(90,5){\line(1,0){10}}
\put(100,0){\makebox(15,10){$
 \scriptstyle  0
$}}
\put(32.5,10){\line(0,1){8}}
\put(25,20){\makebox(15,10){$
  \scriptstyle 2
$}}
\put(82.5,10){\line(0,1){8}}
\put(75,20){\makebox(15,10){$
 \scriptstyle  1
$}}
\end{picture}
                                    \hspace{10pt}
                                        &=&
                                    \hspace{10pt}
\begin{picture}(115,50)(0,5)
%2vertex
\put(0,0){\makebox(15,10)
{$
  \scriptstyle 3
$}}
\put(15,5){\line(1,0){10}}
\put(25,0){\makebox(15,10){$
 \scriptstyle z_1
$}}
\put(40,5){\line(1,0){10}}
\put(50,0){\makebox(15,10)[c]
{$
  \scriptstyle 0
$}}
\put(32.5,10){\line(0,1){8}}
\put(25,20){\makebox(15,10){$
  \scriptstyle 3
$}}
\put(40,25){\line(1,0){10}}
\put(50,20){\makebox(15,10){$
 \scriptstyle z_{21}
$}}
\put(65,25){\line(1,0){10}}
\put(75,20){\makebox(15,10){$
 \scriptstyle  1
$}}
\put(57.5,30){\line(0,1){8}}
\put(50,40){\makebox(15,10){$
 \scriptstyle  2
$}}
\end{picture}
\\
\label{exchange}
\begin{picture}(115,30)(0,5)
%2vertex
\put(0,0){\makebox(15,10)
{$
  \scriptstyle 3
$}}
\put(15,5){\line(1,0){10}}
\put(25,0){\makebox(15,10){$
 \scriptstyle z_2
$}}
\put(40,5){\line(1,0){10}}
\put(50,0){\makebox(15,10)[c]
{$
  \scriptstyle 1
$}}
\put(65,5){\line(1,0){10}}
\put(75,0){\makebox(15,10){$
  \scriptstyle z_1
$}}
\put(90,5){\line(1,0){10}}
\put(100,0){\makebox(15,10){$
 \scriptstyle  0
$}}
\put(32.5,10){\line(0,1){8}}
\put(25,20){\makebox(15,10){$
  \scriptstyle 2
$}}
\put(82.5,10){\line(0,1){8}}
\put(75,20){\makebox(15,10){$
 \scriptstyle  1
$}}
\end{picture}
                                    \hspace{10pt}
                                       &=&\hspace{10pt}\Omega^{\epsilon_{21}}_{321}
                                    \hspace{10pt}
\begin{picture}(115,30)(0,5)
%2vertex
\put(0,0){\makebox(15,10)
{$
  \scriptstyle 3
$}}
\put(15,5){\line(1,0){10}}
\put(25,0){\makebox(15,10){$
 \scriptstyle z_1
$}}
\put(40,5){\line(1,0){10}}
\put(50,0){\makebox(15,10)[c]
{$
  \scriptstyle 2
$}}
\put(65,5){\line(1,0){10}}
\put(75,0){\makebox(15,10){$
  \scriptstyle z_2
$}}
\put(90,5){\line(1,0){10}}
\put(100,0){\makebox(15,10){$
 \scriptstyle  0
$}}
\put(32.5,10){\line(0,1){8}}
\put(25,20){\makebox(15,10){$
  \scriptstyle 1
$}}
\put(82.5,10){\line(0,1){8}}
\put(75,20){\makebox(15,10){$
 \scriptstyle  2
$}}
\end{picture}
\end{eqnarray}
where
$$
\Omega^{\epsilon_{21}}_{321}= {\rm e}^{\epsilon_{21}\pi i(\Delta_3-\Delta_2-\Delta_1)}.
$$
The first of these relations is a direct consequence of (\ref{z_dep}). The second
follows from the fact that $L_{-1}$ acts as the generator of translations (\ref{translation})
\cite{Teschner:2003en}. The third one can be derived as a special limiting case of the braiding
relation by analyzing the analytic continuation of the formula (\ref{B}) from
$
\alpha_1\in {Q\over 2}+i\mathbb{R}
$ to
$\alpha_1=0$
\cite{Ponsot:2001ng,Teschner:2001rv,Teschner:2003en}.

Following \cite{Moore:1988qv,Teschner:2001rv,Teschner:2003en} we define the generalized conformal blocks
in each channel:
\begin{eqnarray*}
{\cal F}^s_{\Delta_s}\!\left[_{\Delta_{4}\;\Delta_{1}}^{\Delta_{3}\;\Delta_{2}}\right]\!(\,z_3,z_2,z_1)
&=&
\begin{picture}(115,30)(0,5)
                                                        %s-block
\put(0,0){\makebox(15,10)
{$
  \scriptstyle 4
$}}
\put(15,5){\line(1,0){10}}
\put(25,0){\makebox(15,10){$
 \scriptstyle z_3
$}}
\put(40,5){\line(1,0){10}}
\put(50,0){\makebox(15,10)[c]
{$
  \scriptstyle s
$}}
\put(65,5){\line(1,0){10}}
\put(75,0){\makebox(15,10){$
  \scriptstyle z_2
$}}
\put(90,5){\line(1,0){10}}
\put(100,0){\makebox(15,10){$
 \scriptstyle  1
$}}
\put(115,5){\line(1,0){10}}
\put(125,0){\makebox(15,10){$
 \scriptstyle z_1
$}}
\put(140,5){\line(1,0){10}}
\put(150,0){\makebox(15,10){$
 \scriptstyle  0
$}}
\put(32.5,10){\line(0,1){8}}
\put(25,20){\makebox(15,10){$
  \scriptstyle 3
$}}
\put(82.5,10){\line(0,1){8}}
\put(75,20){\makebox(15,10){$
 \scriptstyle  2
$}}
\put(132.5,10){\line(0,1){8}}
\put(125,20){\makebox(15,10){$
 \scriptstyle  1
$}}
\end{picture}
\\
{\cal F}^t_{\Delta_t}\!\left[_{\Delta_{4}\;\Delta_{1}}^{\Delta_{3}\;\Delta_{2}}\right]\!(\,z_3,z_2,z_1)
&=&
\begin{picture}(135,50)(0,5)
                                                        %t-block
\put(0,0){\makebox(15,10)
{$
  \scriptstyle 4
$}}
\put(15,5){\line(1,0){10}}
\put(25,0){\makebox(15,10){$
 \scriptstyle z_2
$}}
\put(40,5){\line(1,0){20}}
\put(60,0){\makebox(15,10)[c]
{$
  \scriptstyle 1
$}}
\put(75,5){\line(1,0){20}}
\put(95,0){\makebox(15,10){$
 \scriptstyle z_1
$}}
\put(110,5){\line(1,0){10}}
\put(120,0){\makebox(15,10){$
 \scriptstyle  0
$}}
\put(32.5,10){\line(0,1){8}}
\put(25,20){\makebox(15,10){$
  \scriptstyle t
$}}
\put(40,25){\line(1,0){10}}
\put(50,20){\makebox(15,10){$
\scriptstyle {z_{32}}
$}}
\put(65,25){\line(1,0){10}}
\put(75,20){\makebox(15,10){$
 \scriptstyle  2
$}}
\put(102.5,10){\line(0,1){8}}
\put(95,20){\makebox(15,10){$
 \scriptstyle  1
$}}
\put(57.5,30){\line(0,1){8}}
\put(50,40){\makebox(15,10){$
 \scriptstyle  3
$}}
\end{picture}
\\
{\cal F}^u_{\Delta_u}\!\left[_{\Delta_{4}\;\Delta_{1}}^{\Delta_{3}\;\Delta_{2}}\right]\!(\,z_3,z_2,z_1)
&=&
\begin{picture}(115,30)(0,5)
                                                        %u-block
\put(0,0){\makebox(15,10)
{$
  \scriptstyle 4
$}}
\put(15,5){\line(1,0){10}}
\put(25,0){\makebox(15,10){$
 \scriptstyle z_2
$}}
\put(40,5){\line(1,0){10}}
\put(50,0){\makebox(15,10)[c]
{$
  \scriptstyle u
$}}
\put(65,5){\line(1,0){10}}
\put(75,0){\makebox(15,10){$
 \scriptstyle z_3
$}}
\put(90,5){\line(1,0){10}}
\put(100,0){\makebox(15,10){$
 \scriptstyle  1
$}}
\put(115,5){\line(1,0){10}}
\put(125,0){\makebox(15,10){$
\scriptstyle z_1
$}}
\put(140,5){\line(1,0){10}}
\put(150,0){\makebox(15,10){$
 \scriptstyle  0
$}}
\put(32.5,10){\line(0,1){8}}
\put(25,20){\makebox(15,10){$
  \scriptstyle 2
$}}
\put(82.5,10){\line(0,1){8}}
\put(75,20){\makebox(15,10){$
 \scriptstyle  3
$}}
\put(132.5,10){\line(0,1){8}}
\put(125,20){\makebox(15,10){$
 \scriptstyle  1
$}}
\end{picture}
\end{eqnarray*}
One can easily calculate their relations to the BPZ conformal block (\ref{block})
using (\ref{z_dep}), (\ref{ggg}) and (\ref{simple}):
\begin{eqnarray}
\label{s}
{\cal F}^s_{\Delta_s}\!\left[_{\Delta_{4}\;\Delta_{1}}^{\Delta_{3}\;\Delta_{2}}\right]\!(\,z_3,z_2,z_1)
&=&
(z_3-z_1)^{\Delta_4-\Delta_3-\Delta_2-\Delta_1}
{\cal F}_{\Delta_s}\!\left[_{\Delta_{4}\;\Delta_{1}}^{\Delta_{3}\;\Delta_{2}}\right]\!\left({z_2-z_1\over z_3-z_1}\right),
\\
\label{t}
{\cal F}^t_{\Delta_t}\!\left[_{\Delta_{4}\;\Delta_{1}}^{\Delta_{3}\;\Delta_{2}}\right]\!(\,z_3,z_2,z_1)
&=&
{\rm e}^{-\epsilon_{32}\pi i(\Delta_t-\Delta_3-\Delta_2)}
\nonumber
\\
&\times&
(z_2-z_1)^{\Delta_4-\Delta_3-\Delta_2-\Delta_1}
{\cal F}_{\Delta_t}\!\left[_{\Delta_{4}\;\Delta_{2}}^{\Delta_{1}\;\Delta_{3}}\right]\!\left(1-{z_3-z_1\over z_2-z_1}\right),
\\
\label{u}
{\cal F}^u_{\Delta_u}\!\left[_{\Delta_{4}\;\Delta_{1}}^{\Delta_{3}\;\Delta_{2}}\right]\!(\,z_3,z_2,z_1)
&=&
(z_2-z_1)^{\Delta_4-\Delta_3-\Delta_2-\Delta_1}
{\cal F}_{\Delta_u}\!\left[_{\Delta_{4}\;\Delta_{1}}^{\Delta_{2}\;\Delta_{3}}\right]\!\left({z_3-z_1\over z_2-z_1}\right)
\\
\nonumber
&=&
(z_2-z_1)^{-2\Delta_2}
{\cal F}_{\Delta_u}\!\left[_{\Delta_{1}\;\Delta_{4}}^{\Delta_{3}\;\Delta_{2}}\right]\!\left({z_3-z_1\over z_2-z_1}\right).
\end{eqnarray}

The braiding relation (\ref{braiding}) implies that the braiding matrix (\ref{BE}) can be seen as
the $s-u$ monodromy matrix for generalized conformal blocks
$$
{\cal F}^s_{\Delta_s}\!\left[_{\Delta_{4}\;\Delta_{1}}^{\Delta_{3}\;\Delta_{2}}\right]\!(\,z_3,z_2,z_1) \; = \;
{1\over 2i}\!\!\!\int\limits_{{Q\over 2}+i \mathbb{R}}\!\!\!\! d\alpha_u\;
B^{\epsilon_{32}}_{\alpha_s\,\alpha_u}\!\Big[ \,{_{\alpha_{4}\;\alpha_{1}}^{\alpha_{3}\;\alpha_{2}}} \Big]
{\cal F}^u_{\Delta_u}\!\left[_{\Delta_{4}\;\Delta_{1}}^{\Delta_{3}\;\Delta_{2}}\right]\!(\,z_3,z_2,z_1),
$$
or in the graphic representation
\begin{eqnarray*}
\begin{picture}(165,30)(0,5)
                                                        %s-block
\put(0,0){\makebox(15,10)
{$
  \scriptstyle 4
$}}
\put(15,5){\line(1,0){10}}
\put(25,0){\makebox(15,10){$
 \scriptstyle z_3
$}}
\put(40,5){\line(1,0){10}}
\put(50,0){\makebox(15,10)[c]
{$
  \scriptstyle s
$}}
\put(65,5){\line(1,0){10}}
\put(75,0){\makebox(15,10){$
  \scriptstyle z_2
$}}
\put(90,5){\line(1,0){10}}
\put(100,0){\makebox(15,10){$
 \scriptstyle  1
$}}
\put(115,5){\line(1,0){10}}
\put(125,0){\makebox(15,10){$
 \scriptstyle z_1
$}}
\put(140,5){\line(1,0){10}}
\put(150,0){\makebox(15,10){$
 \scriptstyle  0
$}}
\put(32.5,10){\line(0,1){8}}
\put(25,20){\makebox(15,10){$
  \scriptstyle 3
$}}
\put(82.5,10){\line(0,1){8}}
\put(75,20){\makebox(15,10){$
 \scriptstyle  2
$}}
\put(132.5,10){\line(0,1){8}}
\put(125,20){\makebox(15,10){$
 \scriptstyle  1
$}}
\end{picture}
                                    \hspace{10pt}
                                        &=&
                                        \\
                                       && \hspace{-160pt}
{1\over 2i}\!\!\!\int\limits_{{Q\over 2}+i \mathbb{R}}\!\!\!\! d\alpha_u\;
B^{\epsilon_{32}}_{\alpha_s\,\alpha_u}\!
\Big[ \,{_{\alpha_{4}\;\alpha_{1}}^{\alpha_{3}\;\alpha_{2}}} \Big]
                                    \hspace{5pt}
\begin{picture}(165,30)(0,5)
                                                        %u-block
\put(0,0){\makebox(15,10)
{$
  \scriptstyle 4
$}}
\put(15,5){\line(1,0){10}}
\put(25,0){\makebox(15,10){$
 \scriptstyle z_2
$}}
\put(40,5){\line(1,0){10}}
\put(50,0){\makebox(15,10)[c]
{$
  \scriptstyle u
$}}
\put(65,5){\line(1,0){10}}
\put(75,0){\makebox(15,10){$
 \scriptstyle z_3
$}}
\put(90,5){\line(1,0){10}}
\put(100,0){\makebox(15,10){$
 \scriptstyle  1
$}}
\put(115,5){\line(1,0){10}}
\put(125,0){\makebox(15,10){$
\scriptstyle z_1
$}}
\put(140,5){\line(1,0){10}}
\put(150,0){\makebox(15,10){$
 \scriptstyle  0
$}}
\put(32.5,10){\line(0,1){8}}
\put(25,20){\makebox(15,10){$
  \scriptstyle 2
$}}
\put(82.5,10){\line(0,1){8}}
\put(75,20){\makebox(15,10){$
 \scriptstyle  3
$}}
\put(132.5,10){\line(0,1){8}}
\put(125,20){\makebox(15,10){$
 \scriptstyle  1
$}}
\end{picture}
\end{eqnarray*}
Using (\ref{BE}), (\ref{s}), (\ref{u}) and setting $(\,z_3,z_2,z_1)=(\,1,z,0)$ one gets the relation
(\ref{s-u}).

In a similar way the formula (\ref{euler}) can be derived
from the special limiting  case  of the braiding relation (\ref{exchange})
\begin{eqnarray*}
&&\hspace{-80pt}
\begin{picture}(165,30)(0,5)
                                                        %s-block
\put(0,0){\makebox(15,10)
{$
  \scriptstyle 4
$}}
\put(15,5){\line(1,0){10}}
\put(25,0){\makebox(15,10){$
 \scriptstyle z_3
$}}
\put(40,5){\line(1,0){10}}
\put(50,0){\makebox(15,10)[c]
{$
  \scriptstyle s
$}}
\put(65,5){\line(1,0){10}}
\put(75,0){\makebox(15,10){$
  \scriptstyle z_2
$}}
\put(90,5){\line(1,0){10}}
\put(100,0){\makebox(15,10){$
 \scriptstyle  1
$}}
\put(115,5){\line(1,0){10}}
\put(125,0){\makebox(15,10){$
 \scriptstyle z_1
$}}
\put(140,5){\line(1,0){10}}
\put(150,0){\makebox(15,10){$
 \scriptstyle  0
$}}
\put(32.5,10){\line(0,1){8}}
\put(25,20){\makebox(15,10){$
  \scriptstyle 3
$}}
\put(82.5,10){\line(0,1){8}}
\put(75,20){\makebox(15,10){$
 \scriptstyle  2
$}}
\put(132.5,10){\line(0,1){8}}
\put(125,20){\makebox(15,10){$
 \scriptstyle  1
$}}
\end{picture}\\
&=&                          \hspace{10pt}\Omega^{\epsilon_{21}}_{s21}\hspace{10pt}
\begin{picture}(165,30)(0,5)
                                                        %s-block
\put(0,0){\makebox(15,10)
{$
  \scriptstyle 4
$}}
\put(15,5){\line(1,0){10}}
\put(25,0){\makebox(15,10){$
 \scriptstyle z_3
$}}
\put(40,5){\line(1,0){10}}
\put(50,0){\makebox(15,10)[c]
{$
  \scriptstyle s
$}}
\put(65,5){\line(1,0){10}}
\put(75,0){\makebox(15,10){$
  \scriptstyle z_1
$}}
\put(90,5){\line(1,0){10}}
\put(100,0){\makebox(15,10){$
 \scriptstyle  2
$}}
\put(115,5){\line(1,0){10}}
\put(125,0){\makebox(15,10){$
 \scriptstyle z_2
$}}
\put(140,5){\line(1,0){10}}
\put(150,0){\makebox(15,10){$
 \scriptstyle  0
$}}
\put(32.5,10){\line(0,1){8}}
\put(25,20){\makebox(15,10){$
  \scriptstyle 3
$}}
\put(82.5,10){\line(0,1){8}}
\put(75,20){\makebox(15,10){$
 \scriptstyle  1
$}}
\put(132.5,10){\line(0,1){8}}
\put(125,20){\makebox(15,10){$
 \scriptstyle  2
$}}
\end{picture}
\end{eqnarray*}
by
using
(\ref{s}) and setting $(\,z_3,z_2,z_1)=(\,1,z,0)$.

The fusion matrix can be defined as the $s-t$ monodromy matrix \cite{Teschner:2001rv,Teschner:2003en}
\begin{equation}
\label{s-t_mono}
{\cal F}^s_{\Delta_s}\!\left[_{\Delta_{4}\;\Delta_{1}}^{\Delta_{3}\;\Delta_{2}}\right]\!(\,z_3,z_2,z_1)\; = \;
{1\over 2i}\!\!\!\int\limits_{{Q\over 2}+i \mathbb{R}}\!\!\!\! d\alpha_t\;
F_{\alpha_s\,\alpha_t}\!\Big[ \,{_{\alpha_{4}\;\alpha_{1}}^{\alpha_{3}\;\alpha_{2}}} \Big]
{\cal F}^t_{\Delta_t}\!\left[_{\Delta_{4}\;\Delta_{1}}^{\Delta_{3}\;\Delta_{2}}\right]\!(\,z_3,z_2,z_1)
\;.
\end{equation}
Using formulae (\ref{braiding}), (\ref{identity}) and (\ref{exchange}) one can find
its relation to the braiding matrix \cite{Teschner:2003en,Moore:1988qv}:
\begin{eqnarray*}
&&\hspace{-80pt}
\begin{picture}(165,30)(0,5)
                                                        %s-block
\put(0,0){\makebox(15,10)
{$
  \scriptstyle 4
$}}
\put(15,5){\line(1,0){10}}
\put(25,0){\makebox(15,10){$
 \scriptstyle z_3
$}}
\put(40,5){\line(1,0){10}}
\put(50,0){\makebox(15,10)[c]
{$
  \scriptstyle s
$}}
\put(65,5){\line(1,0){10}}
\put(75,0){\makebox(15,10){$
  \scriptstyle z_2
$}}
\put(90,5){\line(1,0){10}}
\put(100,0){\makebox(15,10){$
 \scriptstyle  1
$}}
\put(115,5){\line(1,0){10}}
\put(125,0){\makebox(15,10){$
 \scriptstyle z_1
$}}
\put(140,5){\line(1,0){10}}
\put(150,0){\makebox(15,10){$
 \scriptstyle  0
$}}
\put(32.5,10){\line(0,1){8}}
\put(25,20){\makebox(15,10){$
  \scriptstyle 3
$}}
\put(82.5,10){\line(0,1){8}}
\put(75,20){\makebox(15,10){$
 \scriptstyle  2
$}}
\put(132.5,10){\line(0,1){8}}
\put(125,20){\makebox(15,10){$
 \scriptstyle  1
$}}
\end{picture}\\
&=&                          \hspace{10pt}\Omega^{\epsilon_{21}}_{s21}\hspace{10pt}
\begin{picture}(165,30)(0,5)
                                                        %s-block
\put(0,0){\makebox(15,10)
{$
  \scriptstyle 4
$}}
\put(15,5){\line(1,0){10}}
\put(25,0){\makebox(15,10){$
 \scriptstyle z_3
$}}
\put(40,5){\line(1,0){10}}
\put(50,0){\makebox(15,10)[c]
{$
  \scriptstyle s
$}}
\put(65,5){\line(1,0){10}}
\put(75,0){\makebox(15,10){$
  \scriptstyle z_1
$}}
\put(90,5){\line(1,0){10}}
\put(100,0){\makebox(15,10){$
 \scriptstyle  2
$}}
\put(115,5){\line(1,0){10}}
\put(125,0){\makebox(15,10){$
 \scriptstyle z_2
$}}
\put(140,5){\line(1,0){10}}
\put(150,0){\makebox(15,10){$
 \scriptstyle  0
$}}
\put(32.5,10){\line(0,1){8}}
\put(25,20){\makebox(15,10){$
  \scriptstyle 3
$}}
\put(82.5,10){\line(0,1){8}}
\put(75,20){\makebox(15,10){$
 \scriptstyle  1
$}}
\put(132.5,10){\line(0,1){8}}
\put(125,20){\makebox(15,10){$
 \scriptstyle  2
$}}
\end{picture}\\
&=&
                \hspace{10pt}
                {1\over 2i}\!\!\!\int\limits_{{Q\over 2}+i \mathbb{R}}\!\!\!\! d\alpha_t\;
                B^{\epsilon_{31}}_{\alpha_s\,\alpha_t}\!
                \Big[ \,{_{\alpha_{4}\;\alpha_{2}}^{\alpha_{3}\;\alpha_{1}}} \Big]
                \Omega^{\epsilon_{21}}_{s21}
                                    \hspace{5pt}
                                    \begin{picture}(165,30)(0,5)
                                                        %s-block
\put(0,0){\makebox(15,10)
{$
  \scriptstyle 4
$}}
\put(15,5){\line(1,0){10}}
\put(25,0){\makebox(15,10){$
 \scriptstyle z_1
$}}
\put(40,5){\line(1,0){10}}
\put(50,0){\makebox(15,10)[c]
{$
  \scriptstyle t
$}}
\put(65,5){\line(1,0){10}}
\put(75,0){\makebox(15,10){$
  \scriptstyle z_3
$}}
\put(90,5){\line(1,0){10}}
\put(100,0){\makebox(15,10){$
 \scriptstyle  2
$}}
\put(115,5){\line(1,0){10}}
\put(125,0){\makebox(15,10){$
 \scriptstyle z_2
$}}
\put(140,5){\line(1,0){10}}
\put(150,0){\makebox(15,10){$
 \scriptstyle  0
$}}
\put(32.5,10){\line(0,1){8}}
\put(25,20){\makebox(15,10){$
  \scriptstyle 1
$}}
\put(82.5,10){\line(0,1){8}}
\put(75,20){\makebox(15,10){$
 \scriptstyle  3
$}}
\put(132.5,10){\line(0,1){8}}
\put(125,20){\makebox(15,10){$
 \scriptstyle  2
$}}
\end{picture}\\
&=&
                \hspace{10pt}
                {1\over 2i}\!\!\!\int\limits_{{Q\over 2}+i \mathbb{R}}\!\!\!\! d\alpha_t\;
                B^{\epsilon_{31}}_{\alpha_s\,\alpha_t}\!
                \Big[ \,{_{\alpha_{4}\;\alpha_{2}}^{\alpha_{3}\;\alpha_{1}}} \Big]
                \Omega^{\epsilon_{21}}_{s21}
                                    \hspace{5pt}
\begin{picture}(135,50)(0,5)
                                                        %t-blocktwisted
\put(0,0){\makebox(15,10)
{$
  \scriptstyle 4
$}}
\put(15,5){\line(1,0){10}}
\put(25,0){\makebox(15,10){$
 \scriptstyle z_1
$}}
\put(40,5){\line(1,0){10}}
\put(50,0){\makebox(15,10)[c]
{$
  \scriptstyle t
$}}
\put(65,5){\line(1,0){10}}
\put(75,0){\makebox(15,10){$
 \scriptstyle z_2
$}}
\put(90,5){\line(1,0){10}}
\put(100,0){\makebox(15,10){$
 \scriptstyle  0
$}}
\put(32.5,10){\line(0,1){8}}
\put(25,20){\makebox(15,10){$
  \scriptstyle 1
$}}
\put(82.5,10){\line(0,1){8}}
\put(75,20){\makebox(15,10){$
 \scriptstyle  t
$}}
\put(90,25){\line(1,0){10}}
\put(100,20){\makebox(15,10){$
\scriptstyle {z_{32}}
$}}
\put(115,25){\line(1,0){10}}
\put(125,20){\makebox(15,10){$
 \scriptstyle  2
$}}
\put(107.5,30){\line(0,1){8}}
\put(100,40){\makebox(15,10){$
 \scriptstyle  3
$}}
\end{picture}\\
&=&
                \hspace{10pt}
                {1\over 2i}\!\!\!\int\limits_{{Q\over 2}+i \mathbb{R}}\!\!\!\! d\alpha_t\;
                \Omega^{\epsilon_{12}}_{41t}\;
                B^{\epsilon_{31}}_{\alpha_s\,\alpha_t}\!
                \Big[ \,{_{\alpha_{4}\;\alpha_{2}}^{\alpha_{3}\;\alpha_{1}}} \Big]
                \Omega^{\epsilon_{21}}_{s21}
                                    \hspace{5pt}
\begin{picture}(135,50)(0,5)
                                                        %t-block
\put(0,0){\makebox(15,10)
{$
  \scriptstyle 4
$}}
\put(15,5){\line(1,0){10}}
\put(25,0){\makebox(15,10){$
 \scriptstyle z_2
$}}
\put(40,5){\line(1,0){20}}
\put(60,0){\makebox(15,10)[c]
{$
  \scriptstyle 1
$}}
\put(75,5){\line(1,0){20}}
\put(95,0){\makebox(15,10){$
 \scriptstyle z_1
$}}
\put(110,5){\line(1,0){10}}
\put(120,0){\makebox(15,10){$
 \scriptstyle  0
$}}
\put(32.5,10){\line(0,1){8}}
\put(25,20){\makebox(15,10){$
  \scriptstyle t
$}}
\put(40,25){\line(1,0){10}}
\put(50,20){\makebox(15,10){$
\scriptstyle {z_{32}}
$}}
\put(65,25){\line(1,0){10}}
\put(75,20){\makebox(15,10){$
 \scriptstyle  2
$}}
\put(102.5,10){\line(0,1){8}}
\put(95,20){\makebox(15,10){$
 \scriptstyle  1
$}}
\put(57.5,30){\line(0,1){8}}
\put(50,40){\makebox(15,10){$
 \scriptstyle  3
$}}
\end{picture}\\
&=&
\hspace{10pt}
                {1\over 2i}\!\!\!\int\limits_{{Q\over 2}+i \mathbb{R}}\!\!\!\! d\alpha_t\;
                F_{\alpha_s\,\alpha_t}\!
                \Big[ \,{_{\alpha_{4}\;\alpha_{1}}^{\alpha_{3}\;\alpha_{2}}} \Big]
                                    \hspace{5pt}
\begin{picture}(135,50)(0,5)
                                                        %t-block
\put(0,0){\makebox(15,10)
{$
  \scriptstyle 4
$}}
\put(15,5){\line(1,0){10}}
\put(25,0){\makebox(15,10){$
 \scriptstyle z_2
$}}
\put(40,5){\line(1,0){20}}
\put(60,0){\makebox(15,10)[c]
{$
  \scriptstyle 1
$}}
\put(75,5){\line(1,0){20}}
\put(95,0){\makebox(15,10){$
 \scriptstyle z_1
$}}
\put(110,5){\line(1,0){10}}
\put(120,0){\makebox(15,10){$
 \scriptstyle  0
$}}
\put(32.5,10){\line(0,1){8}}
\put(25,20){\makebox(15,10){$
  \scriptstyle t
$}}
\put(40,25){\line(1,0){10}}
\put(50,20){\makebox(15,10){$
\scriptstyle {z_{32}}
$}}
\put(65,25){\line(1,0){10}}
\put(75,20){\makebox(15,10){$
 \scriptstyle  2
$}}
\put(102.5,10){\line(0,1){8}}
\put(95,20){\makebox(15,10){$
 \scriptstyle  1
$}}
\put(57.5,30){\line(0,1){8}}
\put(50,40){\makebox(15,10){$
 \scriptstyle  3
$}}
\end{picture}
\end{eqnarray*}
If we exclude the case when $\arg z_1$ lies between $\arg z_2$ and $\arg z_3$ this yields the
relation mentioned in Introduction
\begin{equation}
\label{f-matrix}
F_{\alpha_s\,\alpha_t}\!
\Big[ \,{_{\alpha_{4}\;\alpha_{1}}^{\alpha_{3}\;\alpha_{2}} }\Big]
\;=\;B_{\alpha_s\,\alpha_t}\!
\Big[ \,{_{\alpha_{4}\;\alpha_{2}}^{\alpha_{3}\;\alpha_{1}}} \Big].
\end{equation}
Setting $(\,z_3,z_2,z_1)=(\,1,z,0)$
in  (\ref{s-t_mono}) and
using  (\ref{s}), (\ref{t}), (\ref{f-matrix}) one gets the formula
(\ref{s-t}).

In order to derive the relation (\ref{u-t}) we consider the $u-t$ monodromy matrix defined by
\begin{equation}
\label{u-t_mono}
{\cal F}^u_{\Delta_u}\!\left[_{\Delta_{4}\;\Delta_{1}}^{\Delta_{3}\;\Delta_{2}}\right]\!(\,z_3,z_2,z_1)\; = \;
{1\over 2i}\!\!\!\int\limits_{{Q\over 2}+i \mathbb{R}}\!\!\!\! d\alpha_u\;
A^{\epsilon_{32}}_{\alpha_u\,\alpha_t}\!\Big[ \,{_{\alpha_{4}\;\alpha_{1}}^{\alpha_{3}\;\alpha_{2}}} \Big]
{\cal F}^{\,t}_{\Delta_t}\!\left[_{\Delta_{4}\;\Delta_{1}}^{\Delta_{3}\;\Delta_{2}}\right]\!(\,z_3,z_2,z_1).
\end{equation}
Using formulae (\ref{braiding}), (\ref{identity}), and (\ref{exchange}) one gets:
\begin{eqnarray*}
&&\hspace{-70pt}
\begin{picture}(165,30)(0,5)
                                                        %s-block
\put(0,0){\makebox(15,10)
{$
  \scriptstyle 4
$}}
\put(15,5){\line(1,0){10}}
\put(25,0){\makebox(15,10){$
 \scriptstyle z_2
$}}
\put(40,5){\line(1,0){10}}
\put(50,0){\makebox(15,10)[c]
{$
  \scriptstyle u
$}}
\put(65,5){\line(1,0){10}}
\put(75,0){\makebox(15,10){$
  \scriptstyle z_3
$}}
\put(90,5){\line(1,0){10}}
\put(100,0){\makebox(15,10){$
 \scriptstyle  1
$}}
\put(115,5){\line(1,0){10}}
\put(125,0){\makebox(15,10){$
 \scriptstyle z_1
$}}
\put(140,5){\line(1,0){10}}
\put(150,0){\makebox(15,10){$
 \scriptstyle  0
$}}
\put(32.5,10){\line(0,1){8}}
\put(25,20){\makebox(15,10){$
  \scriptstyle 2
$}}
\put(82.5,10){\line(0,1){8}}
\put(75,20){\makebox(15,10){$
 \scriptstyle  3
$}}
\put(132.5,10){\line(0,1){8}}
\put(125,20){\makebox(15,10){$
 \scriptstyle  1
$}}
\end{picture}\\
&=&                          \hspace{10pt}\Omega^{\epsilon_{31}}_{u321}\hspace{10pt}
\begin{picture}(165,30)(0,5)
                                                        %s-block
\put(0,0){\makebox(15,10)
{$
  \scriptstyle 4
$}}
\put(15,5){\line(1,0){10}}
\put(25,0){\makebox(15,10){$
 \scriptstyle z_2
$}}
\put(40,5){\line(1,0){10}}
\put(50,0){\makebox(15,10)[c]
{$
  \scriptstyle u
$}}
\put(65,5){\line(1,0){10}}
\put(75,0){\makebox(15,10){$
  \scriptstyle z_1
$}}
\put(90,5){\line(1,0){10}}
\put(100,0){\makebox(15,10){$
 \scriptstyle  3
$}}
\put(115,5){\line(1,0){10}}
\put(125,0){\makebox(15,10){$
 \scriptstyle z_3
$}}
\put(140,5){\line(1,0){10}}
\put(150,0){\makebox(15,10){$
 \scriptstyle  0
$}}
\put(32.5,10){\line(0,1){8}}
\put(25,20){\makebox(15,10){$
  \scriptstyle 2
$}}
\put(82.5,10){\line(0,1){8}}
\put(75,20){\makebox(15,10){$
 \scriptstyle  1
$}}
\put(132.5,10){\line(0,1){8}}
\put(125,20){\makebox(15,10){$
 \scriptstyle  3
$}}
\end{picture}\\
&=&
                \hspace{10pt}
                {1\over 2i}\!\!\!\int\limits_{{Q\over 2}+i \mathbb{R}}\!\!\!\! d\alpha_t\;
                B^{\epsilon_{21}}_{\alpha_u\,\alpha_t}\!
                \Big[ \,{_{\alpha_{4}\;\alpha_{3}}^{\alpha_{2}\;\alpha_{1}}} \Big]
                \Omega^{\epsilon_{31}}_{u31}
                                    \hspace{5pt}
                                    \begin{picture}(165,30)(0,5)
                                                        %s-block
\put(0,0){\makebox(15,10)
{$
  \scriptstyle 4
$}}
\put(15,5){\line(1,0){10}}
\put(25,0){\makebox(15,10){$
 \scriptstyle z_1
$}}
\put(40,5){\line(1,0){10}}
\put(50,0){\makebox(15,10)[c]
{$
  \scriptstyle t
$}}
\put(65,5){\line(1,0){10}}
\put(75,0){\makebox(15,10){$
  \scriptstyle z_2
$}}
\put(90,5){\line(1,0){10}}
\put(100,0){\makebox(15,10){$
 \scriptstyle  3
$}}
\put(115,5){\line(1,0){10}}
\put(125,0){\makebox(15,10){$
 \scriptstyle z_3
$}}
\put(140,5){\line(1,0){10}}
\put(150,0){\makebox(15,10){$
 \scriptstyle  0
$}}
\put(32.5,10){\line(0,1){8}}
\put(25,20){\makebox(15,10){$
  \scriptstyle 1
$}}
\put(82.5,10){\line(0,1){8}}
\put(75,20){\makebox(15,10){$
 \scriptstyle  2
$}}
\put(132.5,10){\line(0,1){8}}
\put(125,20){\makebox(15,10){$
 \scriptstyle  3
$}}
\end{picture}\\
&=&
                \hspace{10pt}
                {1\over 2i}\!\!\!\int\limits_{{Q\over 2}+i \mathbb{R}}\!\!\!\! d\alpha_t\;
                \Omega^{\epsilon_{23}}_{t23}\;
                B^{\epsilon_{21}}_{\alpha_u\,\alpha_t}\!
                \Big[ \,{_{\alpha_{4}\;\alpha_{3}}^{\alpha_{2}\;\alpha_{1}}} \Big]
                \Omega^{\epsilon_{31}}_{u31}
                                    \hspace{5pt}
\begin{picture}(165,30)(0,5)
                                                        %s-block
\put(0,0){\makebox(15,10)
{$
  \scriptstyle 4
$}}
\put(15,5){\line(1,0){10}}
\put(25,0){\makebox(15,10){$
 \scriptstyle z_1
$}}
\put(40,5){\line(1,0){10}}
\put(50,0){\makebox(15,10)[c]
{$
  \scriptstyle t
$}}
\put(65,5){\line(1,0){10}}
\put(75,0){\makebox(15,10){$
  \scriptstyle z_3
$}}
\put(90,5){\line(1,0){10}}
\put(100,0){\makebox(15,10){$
 \scriptstyle  2
$}}
\put(115,5){\line(1,0){10}}
\put(125,0){\makebox(15,10){$
 \scriptstyle z_2
$}}
\put(140,5){\line(1,0){10}}
\put(150,0){\makebox(15,10){$
 \scriptstyle  0
$}}
\put(32.5,10){\line(0,1){8}}
\put(25,20){\makebox(15,10){$
  \scriptstyle 1
$}}
\put(82.5,10){\line(0,1){8}}
\put(75,20){\makebox(15,10){$
 \scriptstyle  3
$}}
\put(132.5,10){\line(0,1){8}}
\put(125,20){\makebox(15,10){$
 \scriptstyle  2
$}}
\end{picture}\\
&=&
                \hspace{10pt}
                {1\over 2i}\!\!\!\int\limits_{{Q\over 2}+i \mathbb{R}}\!\!\!\! d\alpha_t\;
                \Omega^{\epsilon_{23}}_{t23}\;
                B^{\epsilon_{21}}_{\alpha_u\,\alpha_t}\!
                \Big[ \,{_{\alpha_{4}\;\alpha_{3}}^{\alpha_{2}\;\alpha_{1}}} \Big]
                \Omega^{\epsilon_{31}}_{u31}
                                    \hspace{5pt}
\begin{picture}(135,50)(0,5)
                                                        %t-blocktwisted
\put(0,0){\makebox(15,10)
{$
  \scriptstyle 4
$}}
\put(15,5){\line(1,0){10}}
\put(25,0){\makebox(15,10){$
 \scriptstyle z_1
$}}
\put(40,5){\line(1,0){10}}
\put(50,0){\makebox(15,10)[c]
{$
  \scriptstyle t
$}}
\put(65,5){\line(1,0){10}}
\put(75,0){\makebox(15,10){$
 \scriptstyle z_2
$}}
\put(90,5){\line(1,0){10}}
\put(100,0){\makebox(15,10){$
 \scriptstyle  0
$}}
\put(32.5,10){\line(0,1){8}}
\put(25,20){\makebox(15,10){$
  \scriptstyle 1
$}}
\put(82.5,10){\line(0,1){8}}
\put(75,20){\makebox(15,10){$
 \scriptstyle  t
$}}
\put(90,25){\line(1,0){10}}
\put(100,20){\makebox(15,10){$
\scriptstyle {z_{32}}
$}}
\put(115,25){\line(1,0){10}}
\put(125,20){\makebox(15,10){$
 \scriptstyle  2
$}}
\put(107.5,30){\line(0,1){8}}
\put(100,40){\makebox(15,10){$
 \scriptstyle  3
$}}
\end{picture}\\
&=&
                \hspace{10pt}
                {1\over 2i}\!\!\!\int\limits_{{Q\over 2}+i \mathbb{R}}\!\!\!\! d\alpha_t\;
                \Omega^{\epsilon_{12}}_{41t}\;
                \Omega^{\epsilon_{23}}_{t23}\;
                B^{\epsilon_{21}}_{\alpha_u\,\alpha_t}\!
                \Big[ \,{_{\alpha_{4}\;\alpha_{3}}^{\alpha_{2}\;\alpha_{1}}} \Big]
                \Omega^{\epsilon_{31}}_{u31}
                                    \hspace{5pt}
\begin{picture}(135,50)(0,5)
                                                        %t-block
\put(0,0){\makebox(15,10)
{$
  \scriptstyle 4
$}}
\put(15,5){\line(1,0){10}}
\put(25,0){\makebox(15,10){$
 \scriptstyle z_2
$}}
\put(40,5){\line(1,0){20}}
\put(60,0){\makebox(15,10)[c]
{$
  \scriptstyle 1
$}}
\put(75,5){\line(1,0){20}}
\put(95,0){\makebox(15,10){$
 \scriptstyle z_1
$}}
\put(110,5){\line(1,0){10}}
\put(120,0){\makebox(15,10){$
 \scriptstyle  0
$}}
\put(32.5,10){\line(0,1){8}}
\put(25,20){\makebox(15,10){$
  \scriptstyle t
$}}
\put(40,25){\line(1,0){10}}
\put(50,20){\makebox(15,10){$
\scriptstyle {z_{32}}
$}}
\put(65,25){\line(1,0){10}}
\put(75,20){\makebox(15,10){$
 \scriptstyle  2
$}}
\put(102.5,10){\line(0,1){8}}
\put(95,20){\makebox(15,10){$
 \scriptstyle  1
$}}
\put(57.5,30){\line(0,1){8}}
\put(50,40){\makebox(15,10){$
 \scriptstyle  3
$}}
\end{picture}\\
&=&
\hspace{10pt}
                {1\over 2i}\!\!\!\int\limits_{{Q\over 2}+i \mathbb{R}}\!\!\!\! d\alpha_t\;
                A^{\epsilon_{32}}_{\alpha_u\,\alpha_t}\!
                \Big[ \,{_{\alpha_{4}\;\alpha_{1}}^{\alpha_{3}\;\alpha_{2}}} \Big]
\hspace{5pt}
\begin{picture}(135,50)(0,5)
                                                        %t-block
\put(0,0){\makebox(15,10)
{$
  \scriptstyle 4
$}}
\put(15,5){\line(1,0){10}}
\put(25,0){\makebox(15,10){$
 \scriptstyle z_2
$}}
\put(40,5){\line(1,0){20}}
\put(60,0){\makebox(15,10)[c]
{$
  \scriptstyle 1
$}}
\put(75,5){\line(1,0){20}}
\put(95,0){\makebox(15,10){$
 \scriptstyle z_1
$}}
\put(110,5){\line(1,0){10}}
\put(120,0){\makebox(15,10){$
 \scriptstyle  0
$}}
\put(32.5,10){\line(0,1){8}}
\put(25,20){\makebox(15,10){$
  \scriptstyle t
$}}
\put(40,25){\line(1,0){10}}
\put(50,20){\makebox(15,10){$
\scriptstyle {z_{32}}
$}}
\put(65,25){\line(1,0){10}}
\put(75,20){\makebox(15,10){$
 \scriptstyle  2
$}}
\put(102.5,10){\line(0,1){8}}
\put(95,20){\makebox(15,10){$
 \scriptstyle  1
$}}
\put(57.5,30){\line(0,1){8}}
\put(50,40){\makebox(15,10){$
 \scriptstyle  3
$}}
\end{picture}
\end{eqnarray*}
With the same restriction for arguments of $z_i$ as in the case of the fusion matrix
this implies
\begin{equation}
\label{a-matrix}
A^{\epsilon_{32}}_{\alpha_u\,\alpha_t}\!
\Big[ \,{_{\alpha_{4}\;\alpha_{1}}^{\alpha_{3}\;\alpha_{2}} }\Big] \;=\;
{\rm e}^{\epsilon_{32}\pi i(\Delta_3+\Delta_2-\Delta_t)}
B_{\alpha_u\,\alpha_t}\!
\Big[ \,{_{\alpha_{4}\;\alpha_{3}}^{\alpha_{2}\;\alpha_{1}}} \Big].
\end{equation}
Setting $(\,z_3,z_2,z_1)=(\,z,1,0)$
in  (\ref{u-t_mono}) and
using  (\ref{t}), (\ref{u}), (\ref{a-matrix}) one gets the formula
(\ref{u-t}).

\section{External weight $\Delta(1,2)$}

If one of the external conformal weights $\Delta_i,\ i=1,\dots,4$ corresponds to a degenerate
Virasoro representation and  the intermediate
weight $\Delta_s$ satisfies an appropriate  fusion rule then the conformal block
is a solution of a certain ordinary differential equation
and the analytic continuation formulae contain only finite number of conformal blocks \cite{Belavin:1984vu}.

In the simplest case of the conformal weight
$$
\delta\; =\; \Delta(-{\textstyle {b\over 2}})\;=\;-{3\over 4}b^2 -{1\over 2}\;=\; \Delta(1,2),
$$
the conformal blocks are solutions of a second order differential equation and can be expressed
in terms of hypergeometric functions \cite{Belavin:1984vu}.
For the weight $\delta$ located at $z_1=0$ one has:
\begin{eqnarray}
\label{block-}
{\cal F}_{\Delta(\alpha_2 -{b\over 2})}
\!\left[_{\Delta_{4}\;\;\delta}^{\Delta_{3}\;\Delta_{2}}\right]
\!\left(z\right)
&=& z^{\Delta(\alpha_2 -{b\over 2}) - \Delta_2-\delta}
(1-z)^{\Delta(\alpha_4 +{b\over 2})- \Delta_2-\Delta_3}
   \\
   \nonumber
    &&\hspace{-50pt} \times\;\;
{}_2F_1\Big(
b(-\alpha_4+ \alpha_3 + \alpha_2  -{\textstyle{b\over 2}} ),\,
b( -\alpha_4 + \bar\alpha_3  + \alpha_2 -{\textstyle{b\over 2}});\,
b(2\alpha_2 -b);\,
z\Big),
 \\[20pt]
 \label{block+}
{\cal F}_{\Delta(\alpha_2 +{b\over 2})}
\!\left[_{\Delta_{4}\;\;\delta}^{\Delta_{3}\;\Delta_{2}}\right]
\!\left(z\right)
&=& z^{\Delta(\alpha_2 +{b\over 2}) - \Delta_2-\delta}
(1-z)^{\Delta(\alpha_4 +{b\over 2} )- \Delta_2-\Delta_3}
\\
\nonumber
    &&\hspace{-50pt} \times\;\;
{}_2F_1\Big(
b(-\alpha_4 + \alpha_3 + \bar\alpha_2  -{\textstyle{b\over 2}} ),\,
b(-\alpha_4  + \bar\alpha_3+ \bar\alpha_2  -{\textstyle{b\over 2}} );\,
b(2\bar\alpha_2 -b);\,
z\Big).
\end{eqnarray}
Using the relation
\begin{equation}
\label{hiperE}
_2F_1(a,b;c;z) = (1-z)^{-a} \;_2F_1\left(a,c-b;c;{z\over z-1}\right)
\end{equation}
valid in the range $|{\arg (1-z)}|<\pi$, and taking into account our choice of cuts
one gets:
\begin{eqnarray}
\label{eulerD}
{\cal F}_{\Delta(\alpha_2 \pm {b\over 2})}
\!\left[_{\Delta_{4}\;\;\delta}^{\Delta_{3}\;\Delta_{2}}\right]
\!\left(z\right) &=&
(1-z)^{\Delta_4-\Delta_3-\Delta_2-\delta}\\
\nonumber
&\times&{\rm e}^{- \epsilon\pi i (\Delta(\alpha_2 \pm {b\over 2})-\Delta_2-\delta)}
{\cal F}_{\Delta(\alpha_2 \pm {b\over 2})}
\!\left[^{\Delta_{3}\;\;\delta}_{\Delta_{4}\;\Delta_{2}}\right]
\!\left({z\over z-1}\right),
\end{eqnarray}
where the conformal blocks with the weight $\delta$ at the location
$z_2=z$ are given by:
\begin{eqnarray*}
{\cal F}_{\Delta(\alpha_1 -{b\over 2})}
\!\left[^{\Delta_{3}\;\;\delta}_{\Delta_{4}\;\Delta_{1}}\right]
\!\left(z\right)
&=&
 z^{\Delta(\alpha_1 -{b\over 2}) - \Delta_1-\delta}
(1-z)^{\Delta(\alpha_3 -{b\over 2})- \Delta_3-\delta}
 \\
&&\hspace{-50pt}\times\;
_2F_1\Big(b(-\alpha_4  +\alpha_3+\alpha_1  -{\textstyle {b\over 2}}),
 b(-\bar\alpha_4+\alpha_3+ \alpha_1 -{\textstyle {b\over 2}});
 b(2\alpha_1 -b);z\Big),
 \\[20pt]
{\cal F}_{\Delta(\alpha_1 +{b\over 2})}
\!\left[^{\Delta_{3}\;\;\delta}_{\Delta_{4}\;\Delta_{1}}\right]
\!\left(z\right)
&=&
z^{\Delta(\alpha_1 +{b\over 2}) - \Delta_1-\delta}
(1-z)^{\Delta(\alpha_3 -{b\over 2})- \Delta_3-\delta}
 \\
&&\hspace{-50pt}\times\;
_2F_1\Big(b(-\alpha_4+\alpha_3  +\bar\alpha_1-{\textstyle {b\over 2}}),
 b(\alpha_4+\alpha_3-\alpha_1  -{\textstyle {b\over 2}});
 b(2\bar\alpha_1 -b);z\Big).
\end{eqnarray*}
Thus in the case under consideration the formula (\ref{euler}) can be seen as a generalization
of the formula (\ref{eulerD}) for the hypergeometric functions.

This is also true for the other formulae. As an example we consider the $s-u$ monodromy (\ref{s-u})
for the blocks (\ref{block-}), (\ref{block+}). It can be easily derived from the relation
\begin{eqnarray*}
F(\alpha,\beta;\gamma;z)
& = &
{
\Gamma(\gamma)\Gamma(\beta-\alpha)
\over
\Gamma(\beta)\Gamma(\gamma-\alpha)
}
(-z)^{-\alpha} F\left(\alpha,1+\alpha-\gamma;1+\alpha - \beta;\frac{1}{z}\right)
\\
&& \\
& + &
{
\Gamma(\gamma)\Gamma(\alpha-\beta)
\over
\Gamma(\alpha)\Gamma(\gamma-\beta)
}
(-z)^{-\beta} F\left(\beta,1+\beta-\gamma;1+\beta-\alpha;\frac{1}{z}\right)
\end{eqnarray*}
valid in the range $|{\arg (-z)}|<\pi$.
One obtains:
\begin{eqnarray}
\label{bb}
{\cal F}_{\Delta_s^\sigma}
\!\left[_{\Delta_{4}\;\;\,\delta}^{\Delta_{3}\;\Delta_{2}}\right]\!\left(z\right)&=&
\\
\nonumber
z^{\Delta_4-\Delta_3-\Delta_2-\delta}\hspace{-5pt}&&\hspace{-20pt}
\sum\limits_{\tau = \pm}
{\rm e}^{-i\pi\epsilon \left(\delta +\Delta_4 -\Delta_s^\sigma
-\Delta_u^\tau \right)} \;{\cal B}_{\,\sigma\,\tau}\;\; {\cal
F}_{\Delta_u^\tau}
\!\left[_{\Delta_{4}\;\;\delta}^{\Delta_{2}\;\Delta_{3}}\right]
\!\left({1\over z} \right)
\\
\nonumber
=\;\;z^{-2\Delta_2}\hspace{-5pt}&&\hspace{-20pt}
\sum\limits_{\tau = \pm}
{\rm e}^{-i\pi\epsilon \left(\delta +\Delta_4 -\Delta_s^\sigma
-\Delta_u^\tau \right)} \;{\cal B}_{\,\sigma\,\tau}\;\; {\cal
F}_{\Delta_u^\tau}
\!\left[_{\;\,\delta\;\;\,\Delta_{4}}^{\Delta_{3}\;\Delta_{2}}\right]
\!\left({1\over z} \right),
\hskip 5mm
\sigma = \pm\ ,
\end{eqnarray}
where $
\Delta_s^\pm =\Delta(\alpha_2\pm{\textstyle {b\over 2}}),\;
\Delta_u^\pm \;=\;\Delta(\alpha_3\pm{\textstyle {b\over 2}}),
$ and
\begin{eqnarray*}
{\cal B}_{--}&=&{
\Gamma(b(2\alpha_2 - b))
\Gamma(b(\bar\alpha_3 - \alpha_3))
\over
\Gamma(b(\bar\alpha_4 - \alpha_3 + \alpha_2 -{\textstyle{b\over 2}}))
\Gamma(b(\alpha_4 - \alpha_3 + \alpha_2 -{\textstyle{b\over 2}}))
}\;,
\\[20pt]
{\cal B}_{-+}&=&
{
\Gamma(b(2\alpha_2 - b))
\Gamma(b(\alpha_3 - \bar\alpha_3))
\over
\Gamma(b(\bar\alpha_4 - \bar\alpha_3 + \alpha_2 -{\textstyle{b\over 2}}))
\Gamma(b(\alpha_4 - \bar\alpha_3 + \alpha_2 -{\textstyle{b\over 2}}))
}\;,
\\[20pt]
{\cal B}_{+-}
&=&
{
\Gamma(b(2\bar\alpha_2 - b))
\Gamma(b(\bar\alpha_3 - \alpha_3))
\over
\Gamma(b(\bar\alpha_4 - \alpha_3 + \bar\alpha_2 -{\textstyle{b\over 2}}))
\Gamma(b(\alpha_4 - \alpha_3 + \bar\alpha_2 -{\textstyle{b\over 2}}))
}\;,
\\[20pt]
{\cal B}_{++}
&=&
{
\Gamma(b(2\bar\alpha_2 - b))
\Gamma(b(\alpha_3 - \bar\alpha_3))
\over
\Gamma(b(\bar\alpha_4 -\bar\alpha_3 +\bar\alpha_2 -{\textstyle{b\over 2}}))
\Gamma(b(\alpha_4 -\bar\alpha_3 + \bar\alpha_2 -{\textstyle{b\over 2}}))
}\;.
\end{eqnarray*}

In order to derive the $s-u$ monodromy
from the general expression (\ref{s-u}) one
needs to analytically continue the integral
\begin{equation}
\label{B:transform}
{1\over 2i}\int\limits_{{Q\over 2} +i \mathbb{R}}\!\!\!d\alpha_u\;
{\rm e}^{-\epsilon\pi i(\Delta_1+\Delta_4-\Delta_s-\Delta_u)}\
B_{\alpha_s\,\alpha_u}\!\Big[
\,{_{\alpha_{4}\;\alpha_{1}}^{\alpha_{3}\;\alpha_{2}}} \Big]\
{\cal F}_{\Delta_u}\!\left[
  _{\Delta_{1}\;\Delta_{4}}^{\Delta_{3}\;\Delta_{2}}\right]\!\left({1\over
    z}\right)
\end{equation}
from the ``physical" values
$\alpha_s,\alpha_1 \in Q/2  + i{\mathbb R}$ to $\alpha_s = \alpha_2 \pm {\textstyle{b\over 2}}$,
 $\alpha_1 = - {\textstyle{b\over 2}}$. Let us note
 that for $\alpha_2,\alpha_3,\alpha_4\in  Q/2  + i{\mathbb R}$
the conformal block in the integrand of (\ref{B:transform}) is regular in this limit.

For $\alpha_1\in  Q/2  + i{\mathbb R}$  the continuation $\alpha_s \to \alpha_2 \pm {\textstyle{b\over 2}}$
of the braiding matrix (\ref{B}) takes the form
\begin{equation}
\label{Bpm}
B_{\alpha_2\pm {\textstyle{b\over 2}},\,\alpha_u}\!\Big[
\,{_{\alpha_{4}\;\alpha_{1}}^{\alpha_{3}\;\alpha_{2}}} \Big]
\; = \;
W_\pm (\alpha_1) U_\pm (\alpha_1) I_\pm (\alpha_1),
\end{equation}
where:
\begin{eqnarray}
\label{W}
W_\pm (\alpha_1)
&=&
{\Gamma_b(2\alpha_2 \pm b)\Gamma_b(2\bar\alpha_2 \mp b)
\over
\Gamma_b(2\bar\alpha_2 - \bar\alpha_1 \mp {\textstyle{b\over 2}})
\Gamma_b(2\bar\alpha_2 - \alpha_1   \mp {\textstyle{b\over 2}})
\Gamma_b(\bar\alpha_1 \pm {\textstyle{b\over 2}})
\Gamma_b(\alpha_1 \pm {\textstyle{b\over 2}})}
\\
\nonumber
& \times&
{1
\over
\Gamma_b(\bar\alpha_4+\bar\alpha_3-\alpha_2 \mp {\textstyle{b\over 2}})
\Gamma_b(\alpha_4+\bar\alpha_3 -\alpha_2 \mp {\textstyle{b\over 2}})
}
\\
\nonumber
& \times&
{1
\over
\Gamma_b(\bar\alpha_4-\alpha_3+ \alpha_2 \pm {\textstyle{b\over 2}})
\Gamma_b(\alpha_4-\alpha_3+\alpha_2 \pm {\textstyle{b\over 2}})}\ ,
\\[15pt]
 \label{U}
 U_\pm (\alpha_1)&=&
{\Gamma_b(\bar\alpha_4+\bar\alpha_2-\alpha_u)
\Gamma_b(\alpha_4+\bar\alpha_2 -\alpha_u)
\Gamma_b(\bar\alpha_4-\alpha_2+ \alpha_u)
\Gamma_b(\alpha_4-\alpha_2+\alpha_u)
}
\\
\nonumber
& \times&
{\Gamma_b(\bar\alpha_1+\bar\alpha_3 -\alpha_u)
\Gamma_b(\alpha_1+\bar\alpha_3-\alpha_u)
\Gamma_b(\bar\alpha_1-\alpha_3+\alpha_u)
\Gamma_b(\alpha_1-\alpha_3+\alpha_u)
\over
\Gamma_b(\bar\alpha_u-\alpha_u)\Gamma_b(\alpha_u-\bar\alpha_u)}\ ,
\\[15pt]
\label{I}
I_\pm(\alpha_1)&=&
{1\over i}\int\limits_{\cal C}\!dt\;
{S_b(\bar\alpha_1+t)
S_b(\alpha_1+t)
S_b(\bar\alpha_4-\alpha_3+\alpha_2+t)
S_b(\alpha_4-\alpha_3+\alpha_2+t)
\over
S_b(Q \mp {\textstyle{b\over 2}} +t)
S_b(2\alpha_2 \pm {\textstyle{b\over 2}}+t)
S_b(\bar\alpha_u+\bar\alpha_3+t)
S_b(\alpha_u+\bar\alpha_3+t)
}\ .
\end{eqnarray}
The contour $\cal C$ in the integral $I_\pm(\alpha_1)$
is deformed such that
 the zeroes of the denominator are located to its right
 and all the  poles of the numerator to its
left.

Let us assume for a moment that the integral $I_\pm(\alpha_1)$ is  regular
in the limit $\alpha_1 \to - {\textstyle{b\over 2}},\ $ i.e. there exists a finite limit
$I_\pm(- {\textstyle{b\over 2}})$ with no poles as a function of $\alpha_u$.
In this case the only poles of the integrand in (\ref{Bpm}) which locations depend on $\alpha_1$
come from the factor
$U_\pm(\alpha_1)$.
One can easily verify that in the limit $\alpha_1 \to - {\textstyle{b\over 2}}$
the contour of integration can be deformed such that
the integral of $U_\pm I_\pm$ is finite.
In this case all the integral vanish due to the
factor $
 \Gamma_b(\alpha_1 \pm {\textstyle{b\over 2}})^{-1}
$ in $W_\pm$ (\ref{W}).

It follows that the only contribution to the  integral (\ref{B:transform}) comes from
the non-regular part of the integral $I_\pm$. Such part can arise only if the contour
$\cal C$ gets ``pinched" between moving poles.
For $\alpha_s=\alpha_2+{b\over 2}$ this happens only for one pair of poles:
the pole  of the factor $S_b(\alpha_1+t)$ at $t=-\alpha_1,$ approaching the contour $\cal C$ from the left,
and the pole of the factor $S_b(Q - {\textstyle{b\over 2}} + t)^{-1}$ at $t={b\over 2},$
located to the right
of $\cal C$.
Moving the contour to the right (or to the left) of this pair (cf. \cite{Ponsot:2000mt}, Lemma 3) one gets:
\begin{eqnarray*}
I_+(\alpha_1) &=& I^0_+(\alpha_1)+ I^R_+(\alpha_1),\\
I^0_+(\alpha_1)
& = &
{S_b(\bar\alpha_1- \alpha_1)
S_b(\bar\alpha_4+\alpha_2-\alpha_3 -\alpha_1)
S_b(\alpha_4+\alpha_2-\alpha_3-\alpha_1)
\over
S_b(Q-{b\over 2}-\alpha_1)
S_b(2\alpha_2+{b\over 2} -\alpha_1)
S_b(\bar\alpha_3+\bar\alpha_u-\alpha_1)
S_b(\bar\alpha_3+ \alpha_u-\alpha_1)}\ ,
\end{eqnarray*}
where $I^R_+(\alpha_1)$ denotes the regular part.

In the case $\alpha_s=\alpha_2-{b\over 2}$ there are two pairs of colliding poles with
the contour $\cal C$ in between and the integral $I_-(\alpha_1)$ can be written as:
\begin{eqnarray*}
I_-(\alpha_1) &=& I^0_-(\alpha_1)+ I^1_-(\alpha_1)+I^R_-(\alpha_1),
\\[15pt]
I^0_-(\alpha_1)
&=&
{S_b(\bar\alpha_1- \alpha_1)
S_b(\bar\alpha_4+\alpha_2-\alpha_3 -\alpha_1)
S_b(\alpha_4+\alpha_2-\alpha_3-\alpha_1)
\over
S_b(Q+{b\over 2}-\alpha_1)
S_b(2\alpha_2-{b\over 2} -\alpha_1)
S_b(\bar\alpha_3+\bar\alpha_u-\alpha_1)
S_b(\bar\alpha_3+ \alpha_u-\alpha_1)}\ ,
\\[15pt]
I^1_-(\alpha_1)& = & -{1\over 2\sin(\pi b^2)}
{S_b(\bar\alpha_1- \alpha_1-b)
\over
S_b(Q-{b\over 2}-\alpha_1)
S_b(2\alpha_2-{b\over 2} -\alpha_1-b)
} \\
& \times &
{
S_b(\bar\alpha_4+\alpha_2-\alpha_3 -\alpha_1-b)
S_b(\alpha_4+\alpha_2-\alpha_3-\alpha_1-b)
\over
S_b(\bar\alpha_3+\bar\alpha_u-\alpha_1-b)
S_b(\bar\alpha_3+ \alpha_u-\alpha_1-b)}\ ,
\end{eqnarray*}
where
$I^0_-(\alpha_1)$ is the contribution from the pole of
 $S_b(\alpha_1+t)$ at $t=-\alpha_1$
and the pole of $S_b(Q + {\textstyle{b\over 2}} + t)^{-1}$ at $t={b\over 2},\ $
$I^1_-(\alpha_1)$ is the contribution from the pole of
 $S_b(\alpha_1+t)$ at $t=-\alpha_1-b$
and the pole of $S_b(Q + {\textstyle{b\over 2}} + t)^{-1}$ at $t=-{b\over 2}$,
and $I^R_-(\alpha_1)$ is the regular part.

$I^0_+(\alpha_1),\ I^0_-(\alpha_1)$ and $I^1_-(\alpha_1)$ are
finite in the limit $\alpha_1\to -{b\over 2}$ and by themselves do not provide
a compensating factor for vanishing $W_\pm$. They however change the structure of
poles in the integrand of  (\ref{B:transform}) which is now  determined by the
factors $U_+I^0_+$,  $U_+I^0_-$, and  $U_+I^1_-$, respectively.

In the case of
$U_+I^0_+$ there are four pairs of poles ``pinching" the contour
of integration. The corresponding contributions can be calculated as in the
case of $t$-integration in terms of residues of the poles at the locations
$\alpha_u= \alpha_3-\alpha_1$, $\bar\alpha_u=\alpha_3-\alpha_1$,
$\alpha_u= \alpha_3+\alpha_1$, $\bar\alpha_u=\alpha_3+\alpha_1$.
Due to the symmetry $\alpha_u\leftrightarrow\bar\alpha_u$ the first two and the last two
residues are equal.
Since
\begin{eqnarray*}
\lim_{\alpha_1\to -{b\over 2}}
W_+(\alpha_1)\;{{1\over i}}\!\!\!\!\!\!\!\!\!
\oint\limits_{\alpha_u =\alpha_3-\alpha_1}\!\!\!\!\!\!\!\!d\alpha_u\,
 U_+(\alpha_1) I^0_+(\alpha_1) &=&
 \\
 &&
 \hspace{-170pt}=\;\;
{
\Gamma(b(2\bar\alpha_2 - b))
\Gamma(b(\alpha_3 - \bar\alpha_3))
\over
\Gamma(b(\bar\alpha_4 -\bar\alpha_3 +\bar\alpha_2 -{\textstyle{b\over 2}}))
\Gamma(b(\alpha_4 -\bar\alpha_3 +\bar\alpha_2 -{\textstyle{b\over 2}}))
}
\;=\;{\cal B}_{\,+\,+}
\\[15pt]
 \lim_{\alpha_1\to -{b\over 2}}
W_+(\alpha_1)\;{{1\over i}}\!\!\!\!\!\!\!\!
\oint\limits_{\alpha_u =\alpha_3+\alpha_1}\!\!\!\!\!\!\!\!d\alpha_u\,
 U_+(\alpha_1) I^0_+(\alpha_1)&=&
 \\
 &&
 \hspace{-170pt}=\;\;
{
\Gamma(b(2\bar\alpha_2 - b))
\Gamma(b(\bar\alpha_3 - \alpha_3))
\over
\Gamma(b(\bar\alpha_4 - \alpha_3 + \bar\alpha_2 -{\textstyle{b\over 2}}))
\Gamma(b(\alpha_4 - \alpha_3 + \bar\alpha_2 -{\textstyle{b\over 2}}))
}
 \;=\;{\cal B}_{\,+\,-}
\end{eqnarray*}
one recovers the formula (\ref{bb}) for $\sigma=+$.

The same  four pairs of colliding poles appear in
 the case of $U_-I^0_-$.
The corresponding residua  are given by:
\begin{eqnarray*}
\lim_{\alpha_1\to -{b\over 2}}
W_-(\alpha_1)\;{{1\over i}}\!\!\!\!\!\!\!\!
\oint\limits_{\alpha_u =\alpha_3-\alpha_1}\!\!\!\!\!\!\!\!d\alpha_u\,
 U_-(\alpha_1) I^0_-(\alpha_1) &=&
 \\
 &&
 \hspace{-170pt}=\;\;
{
\Gamma(b(2\alpha_2-b))
\Gamma(b(\alpha_3-\bar\alpha_3))
\over
\Gamma(b(\bar\alpha_4-\alpha_3+\alpha_2-{b\over 2}))
\Gamma(b(\alpha_4-\alpha_3+ \alpha_2-{b\over 2}))}
\\
 &&
 \hspace{-170pt}\times\;\;
{
\Gamma(b(\bar\alpha_4+\bar\alpha_3-\alpha_2 -{b\over 2})
\Gamma(b(\alpha_4+\bar\alpha_3-\alpha_2 - {b\over 2}))
\over
\Gamma(b(\bar\alpha_4+\alpha_3-\alpha_2-{b\over 2}))
\Gamma(b(\alpha_4+\alpha_3-\alpha_2-{b\over 2}))}
\;\equiv\;{\cal B}^{(0)}_{\,-\,+}
\\[15pt]
 \lim_{\alpha_1\to -{b\over 2}}
W_-(\alpha_1)\;{{1\over i}}\!\!\!\!\!\!\!\!
\oint\limits_{\alpha_u =\alpha_3+\alpha_1}\!\!\!\!\!\!\!\!d\alpha_u\,
 U_-(\alpha_1) I^0_-(\alpha_1)&=&
 \\
 &&
 \hspace{-170pt}=\;\;
{
\Gamma(b(2\alpha_2-b))
\Gamma(b(\bar\alpha_3-\alpha_3))
\over
\Gamma(b(\bar\alpha_4-\alpha_3+\alpha_2-{b\over 2}))
\Gamma(b(\alpha_4-\alpha_3+ \alpha_2-{b\over 2}))}
\;=\;{\cal B}_{\,-\,-}
\end{eqnarray*}
In the case of $U_-I^1_-$ there are only two pairs of colliding poles
contributing the  residua at
$\alpha_u= \alpha_3-\alpha_1$ and $\bar\alpha_u=\alpha_3-\alpha_1$.
By the $\alpha_3\leftrightarrow \bar\alpha_3$ symmetry  they are equal
and take the form
\begin{eqnarray*}
 \lim_{\alpha_1\to -{b\over 2}}
W_-(\alpha_1)\;{{1\over i}}\!\!\!\!\!\!\!\!
\oint\limits_{\alpha_u =\alpha_3-\alpha_1}\!\!\!\!\!\!\!\!d\alpha_u\,
 U_-(\alpha_1) I^1_-(\alpha_1)&=&
 \\
 &&
 \hspace{-170pt}=\;\;
{
\Gamma(b(\bar\alpha_4+\bar\alpha_3-\alpha_2-{b\over 2}))
\Gamma(b(\alpha_4+\bar\alpha_3- \alpha_2-{b\over 2}))
\over
\Gamma(b(\bar\alpha_2-\alpha_2))
\Gamma(b(2\bar\alpha_3-b))}
\;\equiv\;{\cal B}^{(1)}_{\,-\,+}
\end{eqnarray*}
Using properties of the gamma functions and trigonometric
identities one gets:
\begin{eqnarray*}
{\cal B}_{\,-\,+}^{(0)}
+ {\cal B}_{\,-\,+}^{(1)}
& = &
{
\Gamma(b(2\alpha_2 - b))
\Gamma(b(\alpha_3 - \bar\alpha_3))
\over
\Gamma(b(\bar\alpha_4  -\bar\alpha_3 + \alpha_2 - {\textstyle{b\over 2}}))
\Gamma(b(\alpha_4  - \bar\alpha_3 + \alpha_2 - {\textstyle{b\over 2}}))
}
\;\;=\;\;
{\cal B}_{\,-\,+}
\end{eqnarray*}
what agrees with (\ref{bb}) for $\sigma=-$.

\section*{Acknowledgements}

L.H. would like to thank Rainald Flume for numerous helpful
discussions and the faculty of the Physics Institute of the Bonn University
for their hospitality.

\noindent
The work of L.H was partially supported by the DAAD scholarship A/04/24797.

\section*{Appendix A.}
For $\Re\,x > 0$ the Barnes double gamma function has an integral
representation
\[
\log\,\Gamma_b(x)
\; = \;
\int\limits_{0}^{\infty}\frac{dt}{t}
\left[
\frac{{\rm e}^{- x t} - {\rm e}^{- {Q \over 2}t}}
{\left(1-{\rm e}^{- tb}\right)\left(1-{\rm e}^{- t/b}\right)}
-
\frac{\left({\textstyle{Q\over 2}}-x\right)^2}{2{\rm e}^{t}}
-
\frac{{\textstyle{Q\over 2}}-x}{t}
\right].
\]
It satisfies functional relations of the form
\begin{eqnarray*}
\Gamma_b(x+b)
& = &
\frac{\sqrt{2\pi}\,b^{bx-\frac12}}{\Gamma(bx)}\Gamma_b(x),
\\
\Gamma_b\left(x + \frac1b\right)
& = &
\frac{\sqrt{2\pi}\,b^{-\frac{x}{b}+\frac12}}{\Gamma(\frac{x}{b})}\Gamma_b(x),
\end{eqnarray*}
and can be analytically continued to the whole complex $x$ plane as a
meromorphic function with  poles located at $ x = -m b - n{1\over b},\; m, n \in
{\mathbb N}.$

For $x \to 0:$
\[
\Gamma_b(x) \; = \; \frac{\Gamma_b(Q)}{2\pi x} + {\cal O}(1).
\]

For $0 < \Re\, x < Q$ the function $S_b(x) \frac{\Gamma_b(x)}{\Gamma_b(Q-x)}$ can be represented as:
\[
\log\,S_b(x)
\; = \;
\int\limits_{0}^{\infty}
\!\frac{dt}{t}
\left[
\frac{\sinh\left(\frac{Q}{2}-x\right)}{2\sinh\frac{bt}{2}\sinh\frac{t}{2b}}
-\frac{Q/2-x}{t}
\right].
\]
$S_b$ is a meromorphic function of $x$ with poles located at $ x = -m b -
n{1\over b},\;  m, n \in {\mathbb N},$ and zeroes at $ x = Q + m b +
n{1\over b},\;  m, n \in {\mathbb N}.$ It satisfies functional relations of the
form
\begin{eqnarray*}
S_b(x+b)
& = &
2\sin\big(\pi b x\big)\,S_b(x) ,
\\
S_b\left(x + \frac1b\right)
& = &
2\sin\left(\frac{\pi x}{b}\right)S_b(x),
\end{eqnarray*}
and for $x \to 0:$
\[
S_b(x) \; = \; \frac{1}{2\pi x} + {\cal O}(1),
\hskip 2cm
S_b(Q+x) \; = \;
-2\pi x + {\cal O}\left(x^2\right).
\]
Both $\Gamma_b(x)$ and $S_b(x)$ are invariant under $b \to \frac{1}{b}.$
\section*{Appendix B.}
Our aim is to demonstrate the symmetry properties of
the  matrix
$
B_{\alpha_s\,\alpha_u}\!\Big[ \,{_{\alpha_{4}\;\alpha_{1}}^{\alpha_{3}\;\alpha_{2}}} \Big]
$ (\ref{B}).
The conjugations $\alpha_1 \to \bar\alpha_1,$ $\alpha_4 \to
\bar\alpha_4,$ $\alpha_s \to \bar\alpha_s$ and $\alpha_u \to \bar\alpha_u$
are explicit. To see the symmetry of the braiding matrix
under the and exchange of the columns, i.e. simultaneous transformation
$\alpha_1 \leftrightarrow \alpha_4,$ $\alpha_2 \leftrightarrow \alpha_3$
one only needs to shift the integration variable $t \to t + \alpha_3 -
\alpha_2.$

To find the remaining symmetries one can start from the following
identity satisfied by the deformed hypergeometric function
\begin{equation}
\label{deform:identity}
F_b(\alpha,\beta;\gamma;-iy)
\; = \;
{\rm e}^{\pi y(\alpha+\beta-\gamma)}
{
S_b\left(-iy + \frac{Q}{2} - \frac{\alpha + \beta - \gamma}{2}\right)
\over
S_b\left(-iy + \frac{Q}{2} + \frac{\alpha + \beta - \gamma}{2}\right)
}
F_b(\gamma-\alpha,\gamma-\beta;\gamma;-iy).
\end{equation}
From its integral representation
\[
F_b(\alpha,\beta;\gamma;y)
\; = \;
\frac{1}{i}
{S_b(\gamma) \over S_b(\alpha) S_b(\beta)}
\int\limits_{i\mathbb R}\!ds\;
{\rm e}^{2\pi i s y}
{
S_b(\alpha+ s) S_b(\beta + s)
\over
S_b(\gamma+ s) S_b(Q+s)
}
\]
it follows that the integral that appears in the expression for the braiding
matrix,
\[
I
\; = \;
{1\over i}\!
\int\limits_{i\mathbb{R}}\!dt
{S_b(\bar\alpha_1+t)
S_b(\alpha_1+t)
S_b(\bar\alpha_4-\alpha_3+\alpha_2+t)
S_b(\alpha_4-\alpha_3+\alpha_2+t)
\over
S_b(\bar\alpha_u+\bar\alpha_3+t)
S_b(\alpha_u+\bar\alpha_3+t)
S_b(\bar\alpha_s+\alpha_2+t)
S_b(\alpha_s+\alpha_2+t)},
\]
can be expressed as
\begin{eqnarray*}
I & = &
{
S_b(\alpha_4 - \alpha_3 + \alpha_s)
S_b(\bar\alpha_1 - \alpha_2 + \alpha_s)
S_b(\bar\alpha_4 - \bar\alpha_2 + \bar\alpha_u)
S_b(\alpha_1 - \bar\alpha_3 + \bar\alpha_u)
\over
S_b(2\bar\alpha_2)
S_b(2\alpha_3)
}
\\
& \times &
\int\limits_{\mathbb R}\!dy\;
{\rm e}^{2\pi y(\alpha_s + \alpha_3 -\alpha_u  - \alpha_2)}
F_b(
\alpha_4 - \alpha_3 + \alpha_s,
\bar\alpha_1 - \alpha_2 + \alpha_s;
2\bar\alpha_2;-iy)\\
&&
\hskip 3.5cm \times\;
F_b(
\bar\alpha_4 - \bar\alpha_2 + \bar\alpha_u,
\alpha_1 - \bar\alpha_3 + \bar\alpha_u;
2\alpha_3;
iy).
\end{eqnarray*}
Using in this expression the identity (\ref{deform:identity}) we get
\begin{eqnarray}
I & = &
{S_b\left(\alpha_4 + \alpha_s - \alpha_3\right)
S_b\left(Q + \alpha_s - \alpha_2 -\alpha_1\right)
S_b\left(Q - \alpha_4 - \alpha_u + \alpha_2\right)
S_b\left(-\alpha_u + \alpha_3 + \alpha_1\right)
 \over
S_b\left(-\alpha_4 + \alpha_s +\alpha_3\right)
S_b\left(\alpha_s + \alpha_2 + \alpha_1 -Q\right)
S_b\left(Q + \alpha_4 - \alpha_u - \alpha_2\right)
S_b\left(2Q -\alpha_u - \alpha_3 - \alpha_1\right)}
\nonumber \\
&& \nonumber \\
&& \times\;{1\over i}\!
\int\limits_{i\mathbb{R}}\!dt\;
{S_b\left(\alpha_3 + t \right)
S_b\left(\alpha_4 + \alpha_2 + \alpha_1 - Q + t\right)
S_b\left(Q-\alpha_3  + t \right)
S_b\left(\alpha_4 - \alpha_2 + \alpha_1 + t\right)
 \over
S_b\left(\alpha_4 + \alpha_s  + t\right)
S_b\left(Q + \alpha_4  -\alpha_s + t\right)
S_b\left(Q - \alpha_u + \alpha_1 + t\right)
S_b\left(\alpha_u + \alpha_1 + t\right)
},
\nonumber
\end{eqnarray}
what gives
\begin{eqnarray}
\label{B:secondform}
B_{\alpha_s\,\alpha_u}\!\Big[ \,{_{\alpha_{4}\;\alpha_{1}}^{\alpha_{3}\;\alpha_{2}}} \Big]
& = &
{
\Gamma_b(\bar\alpha_2 + \bar\alpha_4 - \alpha_u)
\Gamma_b(\alpha_2 + \bar\alpha_4 - \alpha_u)
\Gamma_b(\bar\alpha_2 - \alpha_4 + \alpha_u)
\Gamma_b(\alpha_2 - \alpha_4 + \alpha_u)
\over
\Gamma_b(\bar\alpha_2 + \alpha_1 - \alpha_s)
\Gamma_b(\alpha_2 + \alpha_1 - \alpha_s)
\Gamma_b(\bar\alpha_2 - \bar\alpha_1 + \alpha_s)
\Gamma_b(\alpha_2 - \bar\alpha_1 + \alpha_s)
}  \nonumber \\
&&  \nonumber \\
& \times &
{
\Gamma_b(\alpha_3 + \alpha_1 - \alpha_u)
\Gamma_b(\bar\alpha_3 + \alpha_1 - \alpha_u)
\Gamma_b(\alpha_3 - \bar\alpha_1 + \alpha_u)
\Gamma_b(\bar\alpha_3 - \bar\alpha_1 + \alpha_u)
\over
\Gamma_b(\alpha_3 + \bar\alpha_4 - \alpha_s)
\Gamma_b(\bar\alpha_3 + \bar\alpha_4 - \alpha_s)
\Gamma_b(\alpha_3 - \alpha_4 + \alpha_s)
\Gamma_b(\bar\alpha_3 - \alpha_4 + \alpha_s)
}
\nonumber\\
&& \nonumber \\
& \times &
{
\Gamma_b(2\alpha_s) \Gamma_b(2Q - 2\alpha_s)
\over
\Gamma_b(Q-2\alpha_u) \Gamma_b(2\alpha_u - Q)
}
\\
&& \nonumber \\
& \times & {1\over i}\!
\int\limits_{i\mathbb{R}}\!dt\;
{
S_b\left(\alpha_3  + t \right)
S_b\left(\bar\alpha_3 + t \right)
S_b\left(\bar\alpha_2 - \bar\alpha_1 + \alpha_4  + t\right)
S_b\left(\alpha_2 - \bar\alpha_1 + \alpha_4  + t\right)
 \over
S_b\left(\bar\alpha_u + \alpha_1 + t\right)
S_b\left(\alpha_u + \alpha_1 + t\right)
S_b\left(\bar\alpha_s  + \alpha_4+ t\right)
S_b\left(\alpha_s  +\alpha_4 + t\right)
} \nonumber \\
& = &
B_{\alpha_s\,\alpha_u}\!\Big[
\,{_{\alpha_{2}\;\bar\alpha_{3}}^{\bar\alpha_{1}\;\alpha_{4}}} \Big].
\nonumber
\end{eqnarray}
This form of the braiding matrix is explicitly invariant under the
conjugations $\alpha_2 \to \bar\alpha_2$ and $\alpha_3 \to \bar\alpha_3.$
Shifting the integration variable $t \to t + \bar\alpha_1 - \alpha_4$
we get
\[
B_{\alpha_s\,\alpha_u}\!\Big[
\,{_{\alpha_{2}\;\bar\alpha_{3}}^{\bar\alpha_{1}\;\alpha_{4}}} \Big]
\; = \;
B_{\alpha_s\,\alpha_u}\!\Big[
\,{_{\bar\alpha_{3}\;\alpha_{2}}^{\alpha_{4}\;\bar\alpha_{1}}} \Big]
\]
and this, together with (\ref{B:secondform}), finally proves the symmetry of
the braiding matrix with respect to the exchange of its rows,
\[
B_{\alpha_s\,\alpha_u}\!\Big[
\,{^{\alpha_{3}\;\alpha_{2}}_{\alpha_{4}\;\alpha_{1}}} \Big]
\; = \;
B_{\alpha_s\,\alpha_u}\!\Big[
\,{_{\alpha_{3}\;\alpha_{2}}^{\alpha_{4}\;\alpha_{1}}} \Big].
\]

\end{document}